\documentclass{pasj01}
\draft 
\Received{$\langle$reception date$\rangle$}
\Accepted{$\langle$acception date$\rangle$}
\Published{$\langle$publication date$\rangle$}
\begin{document}
\title{Cloud-Cloud Collision in the Galactic Center Arc}
\author{Masato Tsuboi$^{1}$,  Yoshimi Kitamura$^1$, Kenta Uehara$^2$, Ryosuke Miyawaki$^3$,  Takahiro Tsutsumi$^4$, Atsushi Miyazaki$^5$,  and Makoto Miyoshi$^6$ }%
\altaffiltext{1}{Institute of Space and Astronautical Science, Japan Aerospace Exploration Agency,\\
3-1-1 Yoshinodai, Chuo-ku, Sagamihara, Kanagawa 252-5210, Japan }
\email{tsuboi@vsop.isas.jaxa.jp}
\altaffiltext{2}{Department of Astronomy, The University of Tokyo, Bunkyo, Tokyo 113-0033, Japan}
\altaffiltext{3}{College of Arts and Sciences, J.F. Oberlin University, Machida, Tokyo 194-0294, Japan}
\altaffiltext{4}{National Radio Astronomy Observatory, P.O.Box O,  Socorro, NM 87801-0387, USA}
\altaffiltext{5}{Japan Space Forum, Kanda-surugadai, Chiyoda-ku,Tokyo,101-0062, Japan}
\altaffiltext{6}{National Astronomical Observatory of Japan, Mitaka, Tokyo 181-8588, Japan}
\KeyWords{Galaxy: center${}_1$ --- stars: formation${}_2$ --- ISM: molecules${}_3$}

\maketitle

\begin{abstract}
We performed a search of  cloud-cloud collision (CCC) sites in the Sagittarius A molecular cloud (SgrAMC) based on the survey observations using the Nobeyama 45-m telescope in the C$^{32}$S $J=1-0$ and SiO $v=0~J=2-1$ emission lines.
 We found candidates being abundant in shocked molecular gas in the Galactic Center Arc (GCA). One of them, M0.014-0.054, is located in the mapping area of our previous ALMA mosaic observation.
We explored the structure and kinematics of M0.014-0.054 in the C$^{32}$S $J=2-1$, C$^{34}$S $J=2-1$, SiO $v=0~J=2-1$, H$^{13}$CO$^+ J=1-0$, and SO $N,J=2,2-1,1$ emission lines and fainter emission lines. 
M0.014-0.054 is likely formed by the CCC between the vertical molecular filaments (VP) of the GCA, and other molecular filaments along Galactic longitude. The bridging features between these colliding filaments on the PV diagram are found, which are the characteristics expected in CCC sites.
We also found continuum compact objects in M0.014-0.054, which have no counterpart in the H42$\alpha$ recombination line. They are detected in the SO emission line, and would be ``Hot Molecular Core (HMC)"s.  
Because the LTE mass of one HMC is larger than the virial mass, it is bound gravitationally.   
This is also detected in the CCS emission line. The embedded star would be too young to ionize the surrounding molecular  cloud.  The VP is traced by poloidal magnetic field. Because the strength of the magnetic field is estimated to be $\sim m$Gauss using the CF method, the VP is supported against fragmentation.
The star formation in the HMC of M0.014-0.054 is likely induced by the CCC between the stable filaments, which may be a common mechanism in the SgrAMC. 
\end{abstract}

\section{Introduction}
The Galactic center region is the nucleus of the nearest spiral galaxy. 
The Central Molecular Zone (CMZ) \citep{MorrisSerabyn} is a large molecular cloud reservoir in the galaxy, which extends along the galactic plane up to $l\sim\pm1^\circ$. The mass of the CMZ is estimated to be  $\sim5\times10^7$ M$_\odot$ (e.g. \cite{MorrisSerabyn};  \cite{Tsuboi1999}). The molecular clouds in the CMZ  are much denser, warmer, and more turbulent than those in the Galactic disk region. 
The CMZ is recognized to be a laboratory for peculiar phenomena, which will be found in central molecular cloud reservoirs of normal external galaxies by future telescopes. 
Young and luminous star clusters which contain over several ten OB stars, for example Arches cluster and  Quintuplet cluster, have been found in the CMZ by IR observations  (e.g. \cite{Genzel}; \cite{Figer1999}; \cite{Figer2002}). They are as luminous as those which are nearly hard to be found in the Galactic disk region.
These star clusters presumably have been formed in the cradle molecular  clouds in the CMZ. The star formation would be influenced by external factors, such as interactions with SNRs and/or cloud-cloud collisions(CCC) because they are crowded in the region (e.g. \cite{Morris1993}; \cite{Hasegawa1994}; \cite{Hasegawa2008}). However, it is difficult to demonstrate observationally how the cradle molecular clouds produce such massive clusters because almost these clusters have already lost the surrounding molecular materials. 
The Galactic Center 50 kms$^{-1}$ molecular cloud (50MC) is an exception, which has still abundant molecular gas and several compact HII regions. In the previous observations, we have found a half-shell structure filled with shocked molecular gas, which would be made by CCC (\cite{Tsuboi2011}, \cite{Tsuboi2015}, \cite{Tsuboi2019}).   As known widely, the dense molecular clouds in the CMZ seem to exist as molecular ridges along the galactic plane, which are bundles of the molecular filaments  (\cite{Bally1987}; \cite{Oka1998}; \cite{Tsuboi1999}). 
The molecular filaments would collide each other,  where star formation would be activated. It is an open issue whether CCCs usually induce star formation in the CMZ or not. 

First, we searched CCC sites based on the survey observations for the Sagittarius A molecular cloud complex (SgrAMC) with the Nobeyama 45-m telescope \citep{Tsuboi1999, Tsuboi2011}. The SgrAMC is one of most conspicuous molecular cloud complexes in the CMZ.  One of the candidates, M0.014-0.054 near the 50MC,  is located serendipitously in the mosaic observation area using ALMA of the 50MC although the observation itself had other science objective \citep{Uehara}. Then we looked for the signs of CCCs and induced star formation in the ALMA data. Throughout this paper, we adopt 8 kpc as the distance to the Galactic center (e.g. \cite{Boehle}). Then, $1\arcsec$ corresponds to about 0.04 pc at the distance. 

\begin{figure}
\begin{center}
\includegraphics[width=18cm, bb=0 0  699.61 592.64]{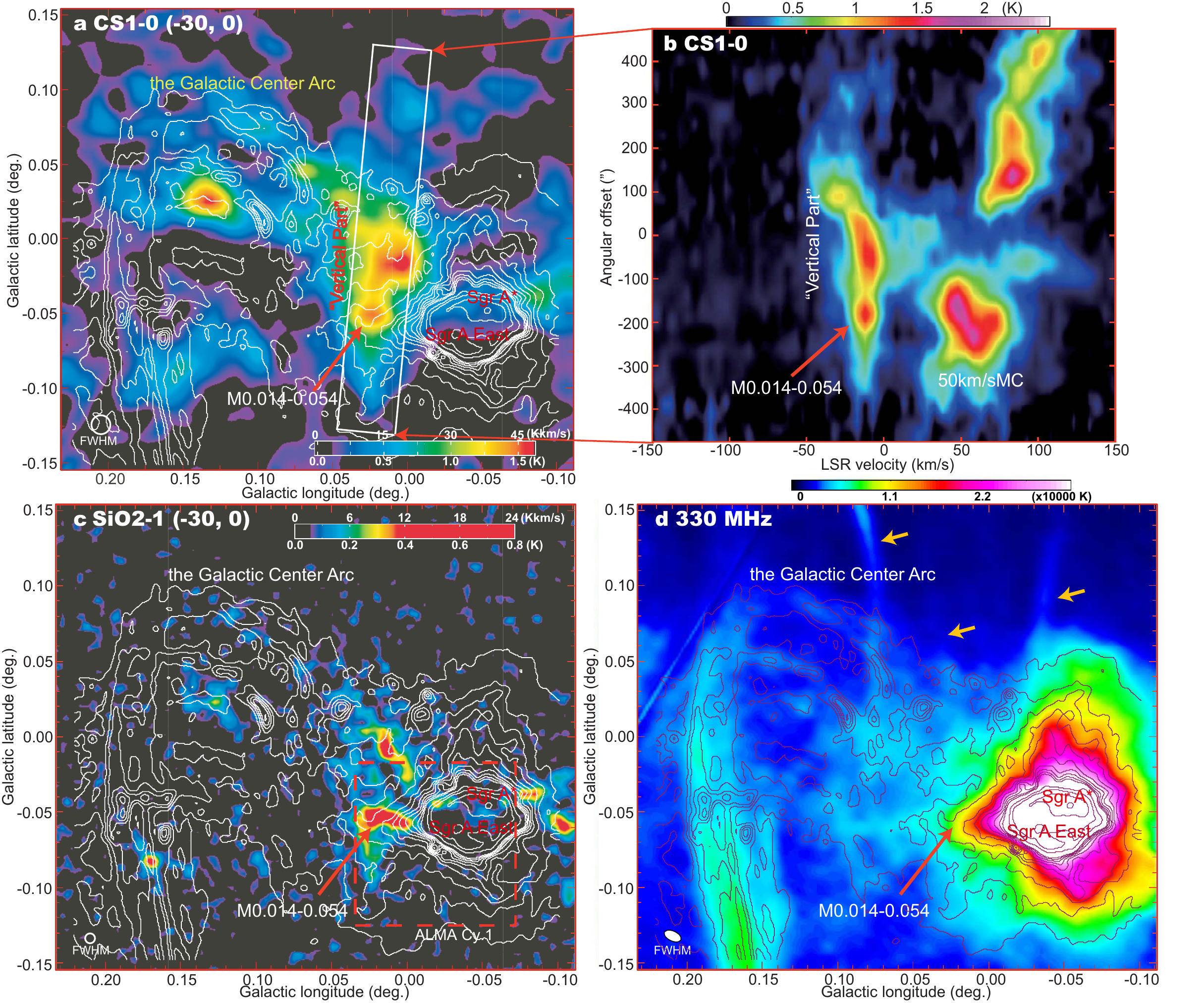}
 \end{center}
 \caption{{\bf a} Integrated intensity map of  the Galactic Center Arc (GCA) in the C$^{32}$S $J=1-0 $ emission line with the Nobeyama 45-m telescope (\cite{Tsuboi1999}).  The velocity range is $V_{\mathrm{LSR}}=-30$ to $0$ km s$^{-1}$. The angular resolution is $45\arcsec$ in FWHM, which indicates as a white circle in the lower left corner. A red arrow indicates a candidate of the CCC site; M0.014-0.054. 
 The contours show the 1.44 GHz (20-cm) continuum emission for comparison (\cite{Yusef-Zadeh1984}).  The contour levels are $(1, 2, 3, 4, 5, 6, 7, 8, 9, 10, 15,  20 ,30, \mathrm{and}~50)\times  95$ K in $T_\mathrm{B}$. The angular resolution is $19\arcsec\times15\arcsec, PA=105^\circ$ in FWHM.
{\bf b}  Position velocity diagram along ``Vertical Part (VP)"  of the GCA extending roughly north and south. The sampling area is shown as a white rectangle in {\bf a}. 
A red arrow indicates M0.014-0.054. {\bf c} Integrated intensity map of  the GCA in the SiO $v=0, J=2-1$ emission line with the Nobeyama 45-m telescope (\cite{Tsuboi2011}).  The velocity range is the same as that of {\bf a}. The angular resolution is $26\arcsec$ in FWHM, which indicates as a white circle in the lower left corner.  
A red arrow indicates M0.014-0.054. The red broken line square  indicates the mapping area of the ALMA Cycle 1 observation (2012.1.00080.S. PI M.Tsuboi).  {\bf d} The 330 MHz (90-cm) continuum emission for comparison \citep{LaRosa}. The angular resolution is $43\arcsec\times24\arcsec, PA=65^\circ$ in FWHM. The yellow arrows show  ``threads".}
 \label{Fig1}
\end{figure}

\section{Search for the Cloud-Cloud Collision Candidates}
We searched the CCC  candidates  in the SgrAMC based on the existing survey observations with the Nobeyama 45-m telescope in the C$^{32}$S $J=1-0$ ($48.990957$ GHz) and SiO $v=0, J=2-1$ ($86.846995$ GHz) emission lines (C$^{32}$S; \cite{Tsuboi1999}, SiO; \cite{Tsuboi2011}).  
These surveys have the highest angular resolutions among single dish observations.
The C$^{32}$S emission line is a tracer of medium dense molecular  cloud, $n({\mathrm H}_2)_{\mathrm{cl}}\sim10^4$ cm$^{-3}$. Because this emission line is moderately optically thick, $\tau\sim0.5-3$, in the SgrAMC (e.g.  \cite{Tsuboi1999}), this emission shows mainly the location of the medium dense molecular cloud. On the other hand, the SiO  emission line is a famous tracer of strong C-shock wave ($\Delta V>30$ km s$^{-1}$) which propagated in the region within $10^5$ yr (e.g. \cite{Gusdorf}, \cite{Jim}).  Because the emission line is often detected in the CCC sites, the detection indicates the CCC candidates.

Figure 1a is the integrated intensity map of  the SgrAMC in the C$^{32}$S $J=1-0 $ emission line with the velocity range of $V_{\mathrm{LSR}}=-30$ to $0$ km s$^{-1}$. The contours in the figure show the Galactic Center Arc (GCA) in the 20-cm continuum emission for comparison (\cite{Yusef-Zadeh1984}).  The molecular cloud is identified as two curved ridges along the GCA (e.g. \cite{Serabyn}). Moreover, a large molecular cloud  connecting with the curved ridges is located apparently in the ``continuum gap" between the GCA and Sagittarius A east SNR. 
This cloud extends vertically north,  $l\sim-0.02^\circ,~b\sim0.05^\circ$, and south, $l\sim0.03^\circ,~b\sim-0.12^\circ$.  We call it  the ``Vertical Part (VP)" here. 
M0.014-0.054 is identified as a strong compact feature which is apparently located on the VP in the C$^{32}$S integrated intensity map. 
Figure 1b is the position-velocity (PV) diagram in the C$^{32}$S $J=1-0 $ emission line along the VP. We identified the VP as a ridge-like feature with no noticeable velocity gradient on the position-velocity diagram. M0.014-0.054 is also seen as a compact feature on the ridge. 

While Figure 1c is the integrated intensity map  in the SiO $v=0, J=2-1$ emission line with  the same area and velocity range as in Figure 1a. Almost all molecular clouds in the area become faint or disappear in the SiO map.   This shows that no strong C-shock wave propagated in the area except for several compact components including M0.014-0.054. 
M0.014-0.054 extends roughly east and west.  The east-west extent of M0.014-0.054 seems to correspond with the width of the VP seen in the C$^{32}$S emission line.   M0.014-0.054 is more prominent in the SiO emission line than in the C$^{32}$S emission line. This shows that shocked molecular gas originated by the C-shock wave  within $10^5$ yr is abundant in M0.014-0.054.  This suggests that the object would be made by some historical event including cloud-cloud collision. 

In addition, Figure 1d shows the continuum emission map at 330 MHz \citep{LaRosa}. Well-known continuum features, ``threads" are identified (yellow arrows). There is also a emission extending from the Sagittarius A east SNR to east, which crosses ``threads" and reaches up to $l\sim0.10^\circ,~b\sim-0.06^\circ$. Because M0.014-0.054 is adjacent to these continuum  features, this would be associated physically with them.

\section{ALMA Observation}   
We have performed the observation of a $330\arcsec \times 330\arcsec$ area covering  the 50MC  in the C$^{32}$S $J=2-1$ ($97.980953$ GHz),  C$^{34}$S $J=2-1$ ($96.412950$ GHz), H$^{13}$CO$^+ J=1-0$ ($86.754288$ GHz), SiO $v=0~J=2-1$ ($86.846995$ GHz), CH$_3$OH ($96.739363, 96.741377,$ and $96.744549$ GHz),  C$_2$H ($87.316925$ and $87.328624$ GHz),  c-C$_3$H$_2$ $J_{K_a, K_c}=2_{1,2}-1_{0,1}$ ($85.338906$ GHz), and SO $N,J=2,2-1,1$ ($86.093983$ GHz)  emission lines and many fainter emission lines  as  ALMA Cycle1 observation (2012.1.00080.S. PI M.Tsuboi). 
We also have observed the continuum emission at 86 GHz, simultaneously.
The entire ALMA observation consists of a 137 pointing mosaic of the 12-m array and a 52 pointing mosaic of the 7-m array (ACA).  Additionally, the single-dish data has been obtained by Total Power Array.
The red broken line square in Figure 1c indicates the mapping area.  It is fortunate that M0.014-0.054 is located in the mapping area of the ALMA observation by chance. 
The molecular emission lines used for imaging here are summarized in Table 1. 
\begin{table}
  \caption{Molecular emission lines used for imaging.  }
  \label{tab:first}
 \begin{center}
    \begin{tabular}{cccc}
 \hline   \hline
Name & transition$^\ast$ &Rest frequency$^\ast$&Remarks\\
& &$\nu_{\mathrm{rest}}$[GHz]&\\
\hline
C$^{32}$S&$J=2-1$&$97.980953$&medium dense molecular cloud\\
&&&moderately optically thick \\
$^{34}$SO&$N,J=2,3-1,2$&$97.715401$&``Hot Molecular Core"\\
CH$_3$OH&$J_{K_a, K_c}=2_{1,1}-1_{1,0}A_{--}$&$97.582808$&mild C-shock \\
CH$_3$OH&$J_{K_a, K_c}=2_{-1,2}-1_{-1,1}E$&$96.739363$&mild C-shock\\
CH$_3$OH&$J_{K_a, K_c}=2_{0,2}-1_{0,1}A_{++}$&$96.741377$&mild C-shock\\
CH$_3$OH&$J_{K_a, K_c}=2_{0,2}-1_{0,1}E$&$96.744549$&mild C-shock\\
CH$_3$CHO&$J_{K_a, K_c}=5_{2,3}-4_{2,2}E$&$96.475523$&\\
C$^{34}$S&$J=2-1$&$96.412950$& dense molecular cloud (opt. thin)\\ C$_2$H&$N=1-0~J=3/2-1/2~F=2-1$&$87.316925$&dense regions exposed to UV radiation$^\dagger$\\
C$_2$H&$N=1-0~J=3/2-1/2~F=1-0$&$87.328624$&dense regions exposed to UV radiation$^\dagger$\\

HN$^{13}$C&$ J=1-0~F=0-1$&$87.090735$&dense molecular cloud (opt. thin)\\
HN$^{13}$C&$ J=1-0~F=2-1$&$87.090859$&dense molecular cloud (opt. thin)\\
HN$^{13}$C&$ J=1-0~F=1-1$&$87.090942$&dense molecular cloud (opt. thin)\\
$^{28}$SiO&$v=0~J=2-1$&$86.846995$&strong C-shock\\
H$^{13}$CO$^+$&$J=1-0$&86.754288&dense molecular cloud (opt. thin)\\
CCS&$N,J=7,6-6,5$&$86.181413$&early stage of star formation\\
$^{32}$SO&$N,J=2,2-1,1$&$86.093983$&``Hot Molecular Core"\\
HCOOH&$J_{K_a, K_c}=4_{1,4}-3_{1,3}$&$86.54618$&\\
HC$^{15}$N&$J=1-0$&$86.054967$&dense molecular cloud (opt. thin)\\
$^{29}$SiO&$v=0~J=2-1$&$85.759188$&strong C-shock\\
HOCO$^+$&$J_{K_a, K_c}=4_{0,4}-3_{0,3}$&$85.531480$&dense molecular cloud (opt. thin)\\
c-C$_3$H$_2$&$J_{K_a, K_c}=2_{1,2}-1_{0,1}$ ortho&$85.338906$&dense regions exposed to UV radiation\\
\hline
Hydrogen atom&H42$\alpha$&$85.6884$&ionized gas\\
\hline
    \end{tabular}
 \end{center}
$^\ast$ https://physics.nist.gov/cgi-bin/micro/table5/start.pl. $^\dagger$ e.g. \cite{Nagy}
\clearpage
\end{table}

The C$^{34}$S and H$^{13}$CO$^+$ emission lines are tracers of dense molecular cloud, $n({\mathrm H}_2)_{\mathrm{cl}}\sim10^5$ cm$^{-3}$. 
These emission lines are thought to be optically thin even in the SgrAMC.  
 The optical thickness of the C$^{34}$S emission line will be demonstrated to be thin for the typical case in the subsection 4.4. That of the H$^{13}$CO$^+$ emission line is estimated to be $\tau<0.2$ based on the observed mean temperature ratio, $T_\mathrm{B}$(H$^{12}$CO$^+$)/$T_\mathrm{B}$( H$^{13}$CO$^+$) $ \sim7$ (e.g. \cite{Armijos2015}).
As mentioned previously, the SiO emission line is a well-known tracer of strong C-shock wave, while the CH$_3$OH molecules are enhanced even by mild C-shock ($\Delta V\sim10$ km s$^{-1}$ e.g. \cite{Hartquist}).  
The C$_2$H and c-C$_3$H$_2$ emission lines are well-known tracers of photodissociation regions (PDRs). The abundance of both radicals was found to increase close
to the dissociation front (e.g. \cite{Jansen}; \cite{Fuente}). 
The HN$^{13}$C and HC$^{15}$N emission lines have critical densities in the environment of the SgrAMC as high as $n({\mathrm H}_2)_{\mathrm{cl}}\sim10^6$ cm$^{-3}$ and $n({\mathrm H}_2)_{\mathrm{cl}}\sim10^7$ cm$^{-3}$, respectively. These intensities can vary widely by their path length, abundance, and isotope ratio.
The SO and $^{34}$SO emission lines are usually emitted from ``Hot Molecular Core (HMC)"s where the molecular cloud is heated up to 100 K by newborn stars. 
The H42$\alpha$ recombination line and continuum emission at 86 GHz are simultaneously observed. The H42$\alpha$ recombination line traces ionized gas. The continuum emission at 86 GHz is usually emitted by ionized gas. However,  there is a possibility that the emission is originated by warm dust or non-thermal mechanism in the Galactic center region. In addition, the CCS emission line is partly in the observation frequency range. The CCS molecule is abundant in the early stage of star formation but decreases with increasing time (e.g. \cite{Hirahara}).

The resultant maps have FWHM angular resolutions of $\sim2\farcs5\times \sim1\farcs8, PA\sim-30^\circ$  and $\sim1\farcs9\times \sim1\farcs3, PA\sim-37^\circ$ using ``natural weighting" and ``briggs weighting" as {\it u-v} sampling, respectively. The original velocity resolution is $1.7$ km s$^{-1}$(488 kHz).
The typical rms noise levels in the emission free areas of the resultant maps using ``natural weighting" and ``briggs weighting" are $\sim0.003$ Jy/beam/$1.7$ km s$^{-1}$ or $\sim0.11$ K in $T_\mathrm{B}$ and $\sim0.005$ Jy/beam/$1.7$ km s$^{-1}$ or $\sim0.17$ K in $T_\mathrm{B}$, respectively.
J0006-0623, J1517-2422, J717-3342,  J1733-1304, J1743-3058, J1744-3116 and J2148+0657 were used as phase calibrators. The flux density scale was determined using Titan, Neptune and Mars.  The calibration and imaging of the data were done by CASA \citep{McMullin}. 

Although the area around M0.014-0.054 is less dense comparing to the 50MC \citep{Uehara}, the C$^{32}$S and C$^{34}$S emission lines are significantly resolved out only by the interferometer observations when they have been processed in the same procedure mentioned above. This is because the molecular cloud is widely extended in the area. 
The combining with the single-dish data, Total Power Array data, is required to recover the missing flux.  We  performed the procedure  using the CASA task "FEATHER". 
We present  the detailed description of the procedure and the full results in other papers (\cite{Uehara}, \cite{Ueharab}).

\section{Results}  
\begin{figure}
\begin{center}
\includegraphics[width=17cm, bb=0 0  558.39 631.61]{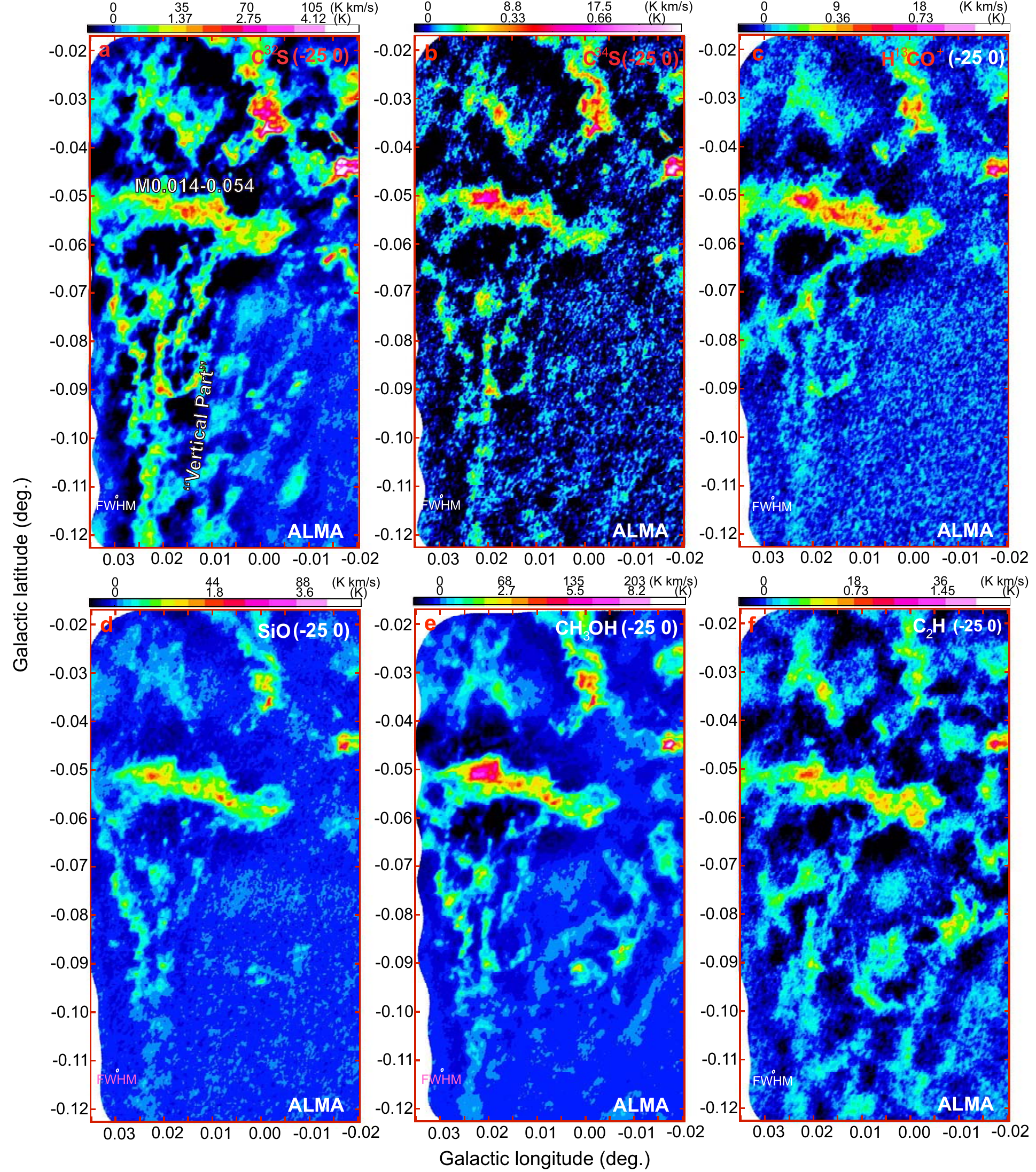}
 \end{center}
 \caption{Integrated intensity maps of M0.014-0.054 and the ``Vertical Part"  in the C$^{32}$S $J=2-1$({\bf a}), C$^{34}$S $J=2-1$({\bf b}), H$^{13}$CO$^+ J=1-0$({\bf c}), SiO $v=0~J=2-1$ ({\bf d}),   CH$_3$OH (96.741 GHz)({\bf e}),  C$_2$H (87.317 GHz)({\bf f}),  c-C$_3$H$_2$ ({\bf g}), HN$^{13}$C ({\bf h}),  HC$^{15}$N ({\bf i}), SO $N,J=2,2-1,1$ ({\bf j}),  and H42$\alpha$({\bf k})  emission lines with ALMA. The C$^{32}$S map is before the combining with the single-dish data. The velocity range is $V_{\mathrm{LSR}}=-25$ to $0$ km s$^{-1}$. The angular resolutions are $\sim2\farcs5\times \sim1\farcs8$ in FWHM and $PA\sim-30^\circ$, which indicate as white ovals in the lower left corners. The rms noise of these maps is $\sim0.036$ K or $0.9$ K km s$^{-1}$. 
The panel ({\bf l}) shows the map of 850 $\mu$m continuum with JCMT is also shown  for comparison (pseudo color: \cite{Pierce-Price2000}). The overlaid contours show the continuum emission at 86 GHz with ALMA. The contour levels are $(1, 2, 4, 8, 16) \times 0.023$ K. The angular resolution  of JCMT is $\sim17\arcsec$ in FWHM.  }
\label{Fig2}
\end{figure}

\begin{figure}
\begin{center}
\setcounter{figure}{1}
\includegraphics[width=17cm, bb=0 0  558.38 632.89]{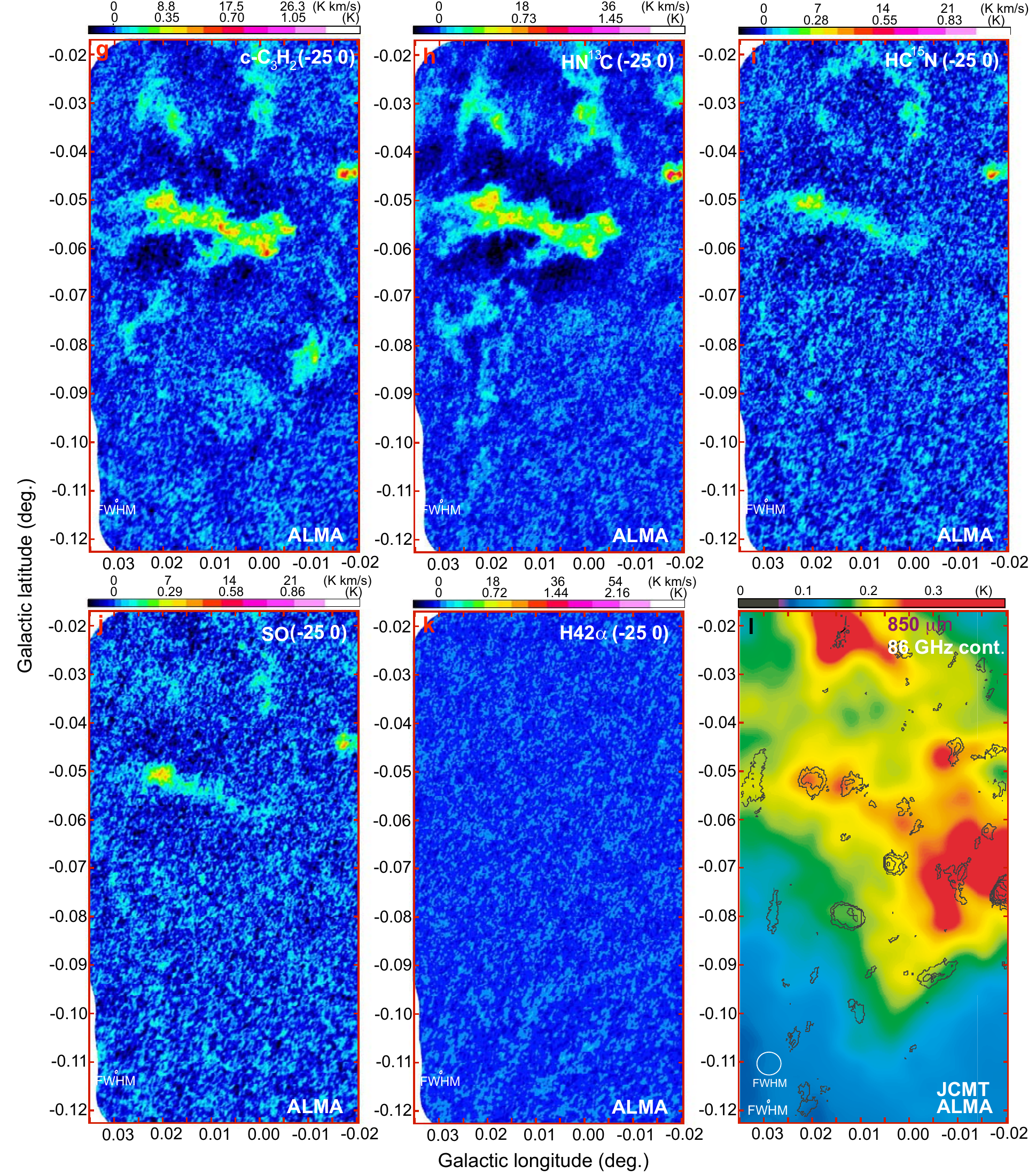}
 \end{center}
 \caption{Continued. 
 }
\label{Fig2}
\end{figure}

\subsection{Distributions of Molecular Line Emissions around M0.014-0.054 and the ``Vertical Part"} 
Figures 2a, 2b, 2c, 2d, and 2e show the integrated intensity maps with ALMA of  M0.014-0.054 and the VP  in the C$^{32}$S,  C$^{34}$S,  H$^{13}$CO$^+$, SiO, and CH$_3$OH (blended lines around 96.741 GHz) emission lines, respectively. 
Note that the data for the C$^{32}$S and C$^{34}$S maps are not combined with the single-dish data. 
The velocity range of each map is $V_{\mathrm{LSR}}=-25$ to $0$ km s$^{-1}$.  
The velocity range was determined from Figures 1b and 1c.

M0.014-0.054 and the VP are clearly detected in the map of the C$^{32}$S emission line. The  VP  is resolved into the bundles of the molecular filaments extending roughly from north to south.  M0.014-0.054 is extended roughly from west to east and over the filaments. The filaments line up nearly at a right angle to M0.014-0.054 in the far distance area. However, they become disordered and tangled in the vicinity of  M0.014-0.054.  The positions of the peaks in the C$^{32}$S emission line do not correspond to those in the maps of the other emission lines which will be mentioned in the following part. This is probably because the C$^{32}$S emission line is partially optically thick in the region. 
M0.014-0.054 is also clearly detected in the maps of the C$^{34}$S and H$^{13}$CO$^+$ emission lines.
They resemble each other very well. These show that  dense molecular cloud with $n({\mathrm H}_2)_{\mathrm{cl}}\sim10^5$ cm$^{-3}$ is abundant in M0.014-0.054.
In contrast, the  molecular filaments in the C$^{34}$S and H$^{13}$CO$^+$ emission lines are fainter than those in the C$^{32}$S emission line, suggesting that this shows that the molecular cloud in the filaments is less dense than that in  M0.014-0.054.

M0.014-0.054 is clearly detected in the maps of the SiO and CH$_3$OH emission lines. The detection in the SiO emission line shows that a strong C-shock wave ($\Delta V>30$ km s$^{-1}$) propagated within $10^5$ yr over the entire M0.014-0.054.
Although the molecular filaments of the VP are also detected in the maps of these emission lines, they are faint on the whole. For example, the prominent parts are located around $l\sim0^\circ.010,~b\sim-0^\circ.060$,  $l\sim0^\circ.015,~b\sim-0^\circ.087$, and $l\sim0^\circ.025,~b\sim-0^\circ.035$ in the maps of the C$^{32}$S and H$^{13}$CO$^+$ emission lines. However, they almost disappear in the SiO emission line. 

The C$_2$H (87.317 GHz) and c-C$_3$H$_2$  emission lines show a similar distribution in M0.014-0.054  (see Figures 2f and 2g). 
The other C$_2$H (87.329 GHz) emission line is also detected although this line is fainter.  The VP almost disappears in the c-C$_3$H$_2$  emission line.
It has been reported  that the $N$[C$_2$H]/$N$[c-C$_3$H$_2$] column density ratio increases
by the UV radiation ($\sim32$ at the PDR edge; \cite{Cuadrado}, up to $\sim80$ at planetary nebulae;  \cite{Schmidt}). Because the VP is less dense than M0.014-0.054, even the interior of the VP should have the high ratio.
The  c-C$_3$H$_2$  and HN$^{13}$C  emission lines show a similar distribution in M0.014-0.054 but the latter is faint (see Figures 2g and 2h). The HC$^{15}$N and SO emission lines show a similar distribution (see Figures 2i and 2j).
M0.014-0.054 is identified as a faint extended feature with a single  compact peak in the maps of the HC$^{15}$N and SO emission lines. The peak of the feature is located in the vicinity of strongest peaks in the other emission lines except for the C$^{32}$S emission line. The filaments of the VP  disappears in the HC$^{15}$N and SO emission lines.

The structures mentioned above including M0.014-0.054 and the filaments of the VP have no ionized gas counterparts which appears in the H42$\alpha$ recombination line (see Figure 2k). However, two distinct components in M0.014-0.054, are identified in the continuum emission at 86 GHz (see Figure 2l).  The east one corresponds to the peak in the SO emission line.  As mentioned previously, the SO emission line is a tracer of HMCs. Then the peaks would involve HMCs. 
The map of  the 850 $\mu$m continuum with JCMT \citep{Pierce-Price2000} is shown for comparison. The whole of M0.014-0.054 is also detected in the 850 $\mu$m continuum although the VP is not identified. Two peaks detected at 850 $\mu$m are associated with the peaks detected at 86 GHz.

\begin{figure}
\begin{center}
\includegraphics[width=18cm, bb=0 0 545.76 280.02]{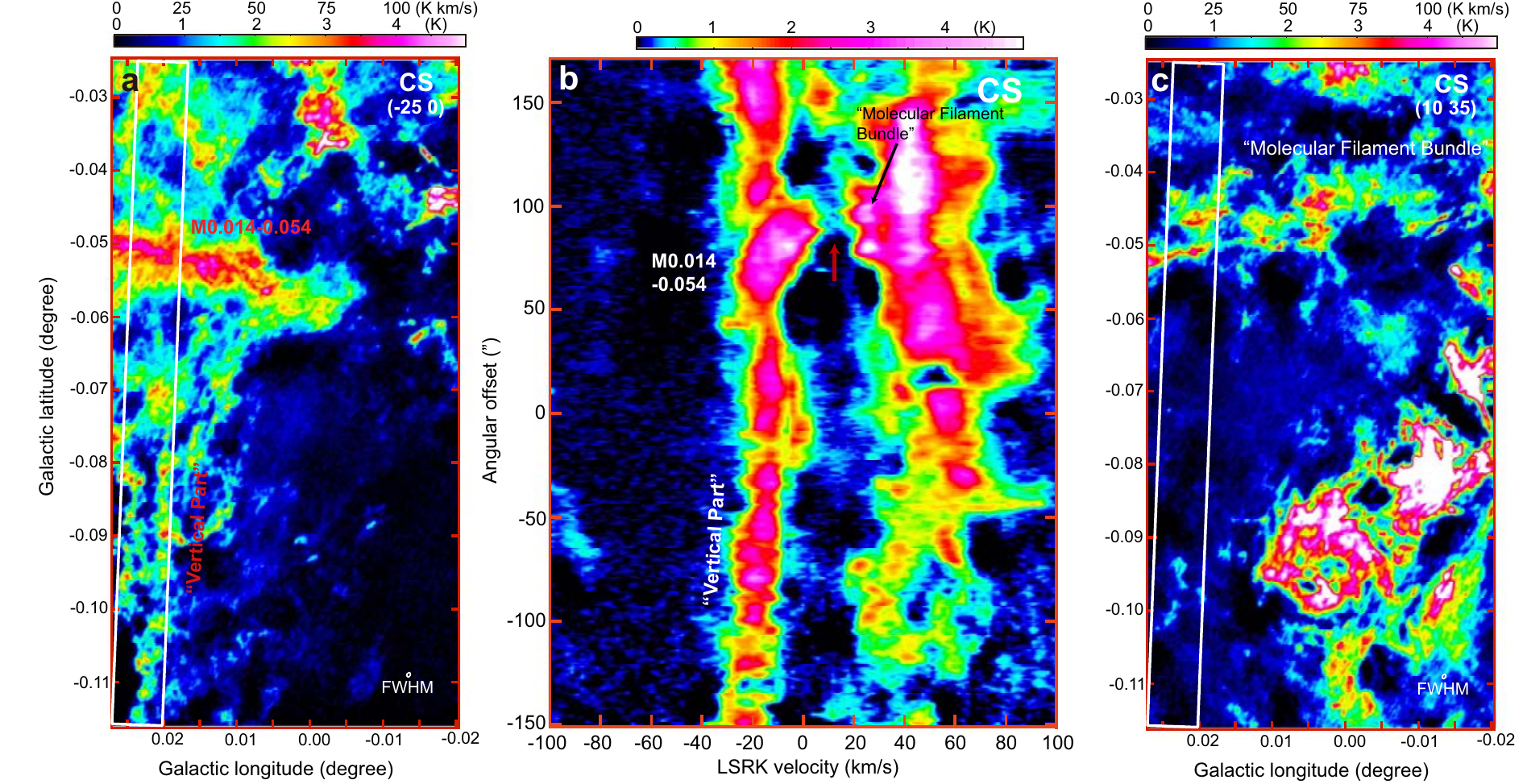}
 \end{center}
 \caption{{\bf a} Integrated intensity map with $V_{\mathrm{LSR}}=-25$ to $0$ km s$^{-1}$ in the C$^{32}$S $J=2-1$ emission line. This is after the combining with the single-dish data. The angular resolution is $\sim1\farcs9\times \sim1\farcs3$ in FWHM and $PA\sim-37^\circ$, which indicates as an oval in the lower right corner. The rectangle shows the sampling area of {\bf b}. {\bf b} Position-velocity diagram  in the C$^{32}$S $J=2-1$ emission line along the ``Vertical Part" (white rectangle in {\bf a} and {\bf c}). {\bf c} Integrated intensity map with $V_{\mathrm{LSR}}=10$ to $35$ km s$^{-1}$ in the C$^{32}$S $J=2-1$ emission line.   }
 \label{Fig3}
\end{figure}
\begin{figure}
\begin{center}
\includegraphics[width=18cm, bb=0 0  555.4 278.05]{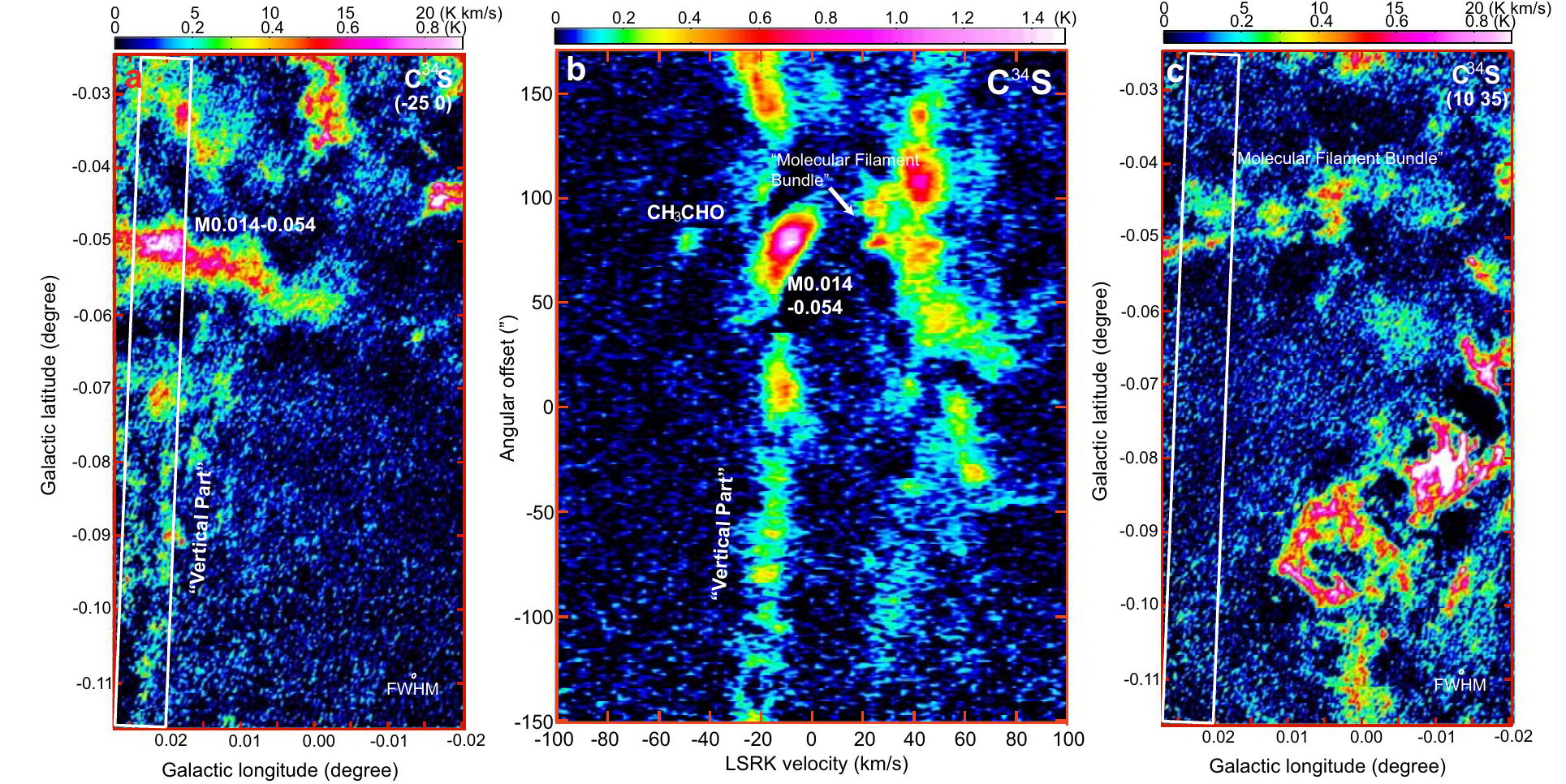}
 \end{center}
 \caption{{\bf a} Integrated intensity map with $V_{\mathrm{LSR}}=-25$ to $0$ km s$^{-1}$ in the C$^{34}$S $J=2-1$ emission line. This is after the combining with the single-dish data. The angular resolution is $\sim1\farcs9\times \sim1\farcs3$ in FWHM and $PA\sim-37^\circ$, which indicates as an oval in the lower right corner. The rectangle shows the sampling area of {\bf b}. {\bf b} Position-velocity diagram  in the C$^{34}$S $J=2-1$ emission line along the ``Vertical Part". {\bf c} Integrated intensity map with $V_{\mathrm{LSR}}=10$ to $35$ km s$^{-1}$ in the C$^{34}$S $J=2-1$ emission line.   }
 \label{Fig4}
\end{figure}

\subsection{Kinematics along the ``Vertical Part"} 
Figures 3a and 4a show the integrated intensity maps with $V_{\mathrm{LSR}}=-25$ to $0$ km s$^{-1}$ in the C$^{32}$S $J=2-1$ and C$^{34}$S $J=2-1$ emission lines, respectively.  
Because the data of these maps are combined with the single-dish data, the mapping areas become narrow slightly. 
The appearances of the VP in these emission lines resemble each other very well. 
Figures 3b and 4b show the PV diagrams along the VP in the C$^{32}$S $J=2-1$ and C$^{34}$S $J=2-1$ emission lines, respectively.   The sampling areas are shown as the rectangles in Figures 3a and 4a. In the PV diagrams, the VP is identified as a long feature with $V_{\mathrm{LSR}}\sim-30$ to $0$ km s$^{-1}$. There is no clear velocity gradient in the feature. The appearance and  velocity gradient are consistent with those of the corresponding feature shown in Figure 1b.  
On the other hand, M0.014-0.054 is identified as a compact curved feature on the filaments of the VP  (Figures 3b and 4b). This feature suggests that M0.014-0.054 had been affected by any external interaction (e.g. \cite{Haworth}, \cite{Haworthb}). 
Additionally,  a large extended feature with $V_{\mathrm{LSR}}\sim30$ to $80$ km s$^{-1}$ is also identified. The appearances and  kinematics of these features are consistent with those of the corresponding features shown in Figure 1b.

In the PV diagrams, there are two compact components with $V_{\mathrm{LSR}}\sim20$ km s$^{-1}$ adjoining the extended feature mentioned above.
In order to clarify the angular extension of these components, the integrated intensity maps with $V_{\mathrm{LSR}}=10$ to $35$ km s$^{-1}$ are shown in Figures 3c and 4c.  These features are resolved into two molecular filaments which are almost parallel to each other in the maps.  
The two components in the PV diagrams are thought to be a part of the molecular filament bundle, which will be discussed in the following subsection.
 
In addition, a faint feature is also identified around the  angular offset of M0.014-0.054 and velocity of $V_{\mathrm{LSR}}\sim-50$ km s$^{-1}$ in the C$^{34}$S $J=2-1$ emission line (see Figure 4b). The feature is the contamination by the CH$_3$CHO  $J_{K_a, K_c}=5_{-2,4}-4_{-2,3}E$ emission line ($96.425620$ GHz) because this feature is not detected in Figure 3b.

\begin{figure}
\begin{center}
\includegraphics[width=17cm, bb=0 0  272.94 306.78]{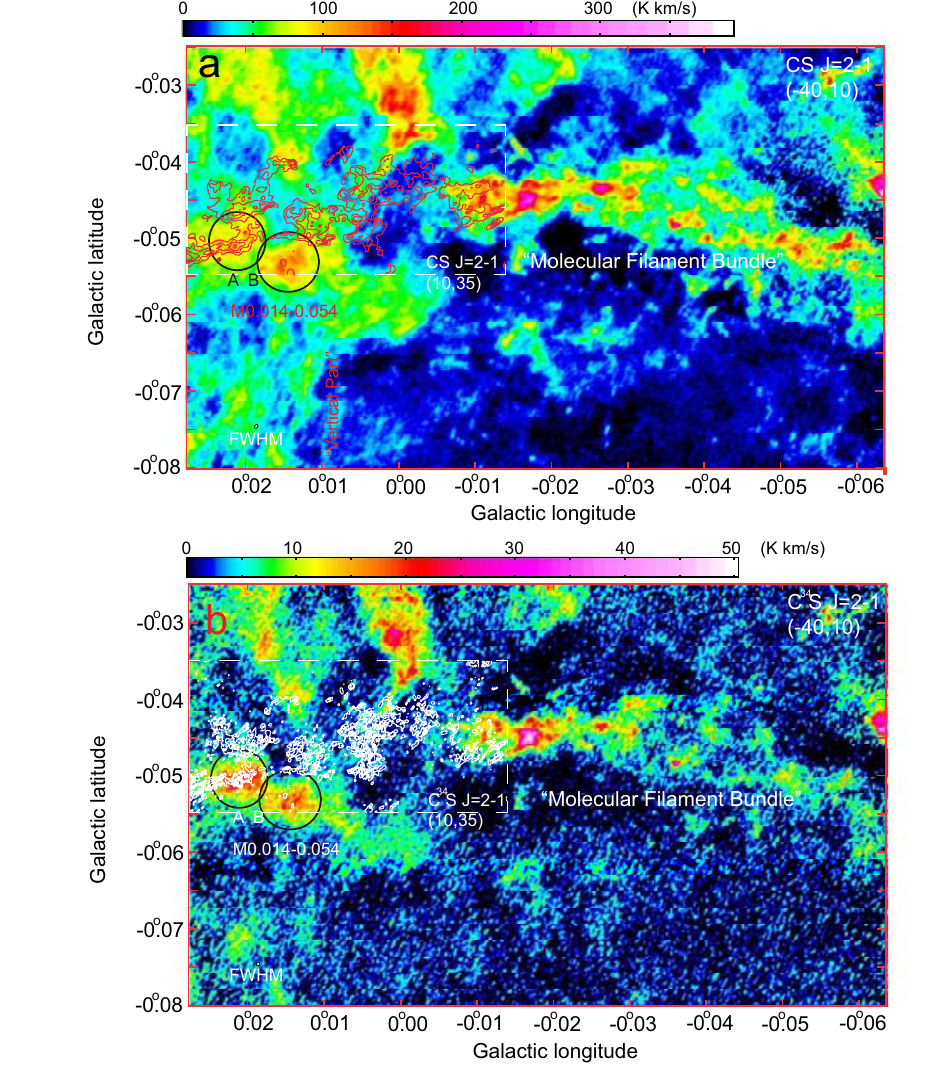}
 \end{center}
 \caption{{\bf a} Integrated intensity maps around the "Molecular Filament Bundle" with $V_{\mathrm{LSR}}=-40$ to $10$ km s$^{-1}$ (pseudo color) and $V_{\mathrm{LSR}}=10$ to $35$ km s$^{-1}$ (contours)  in the C$^{32}$S $J=2-1$ emission line. These are after the combining with the single-dish data. The angular resolution is $\sim1\farcs9\times \sim1\farcs3$ in FWHM and $PA\sim-37^\circ$, which indicates as an oval in the lower left corner. The contour levels are 32,48, 64, and 80 K km s$^{-1}$.
  {\bf b} Integrated intensity maps around the "Molecular Filaments Bundle" with $V_{\mathrm{LSR}}=-40$ to $10$ km s$^{-1}$ (pseudo color) and $V_{\mathrm{LSR}}=10$ to $35$ km s$^{-1}$ (contours)  in the C$^{34}$S $J=2-1$ emission line. These are after the combining with the single-dish data. The angular resolution is $\sim1\farcs9\times \sim1\farcs3$ in FWHM and $PA\sim-37^\circ$, which indicates as an oval in the lower left corner. The contour levels are 4, 6, 8, and 10 K km s$^{-1}$. }
 \label{Fig5}
\end{figure}

\subsection{``Molecular Filament Bundle"}
Another molecular cloud ridge extends roughly east and west over this area (see Figures 5a and 5b). The western and eastern parts of this structure are seen in the negative and positive{LSR} velocities, respectively. We call this structure ``Molecular Filament Bundle" (MFB).
Figure 5a and 5b show the integrated intensity maps around the MFB with $V_{\mathrm{LSR}}=-40$ to $10$ km s$^{-1}$ (pseudo color) and $V_{\mathrm{LSR}}=10$ to $35$ km s$^{-1}$ (contours)  in the C$^{32}$S $J=2-1$ and C$^{34}$S $J=2-1$ emission lines, respectively. The MFB would be extended over the mapping area. 
The MFB is almost along the Galactic longitude but curved slightly.  The MFB is resolved into the bundle of a few thin molecular filaments in these maps. 
The width of the MFB is changed periodically from $\sim40\arcsec$ around $l\sim-0^\circ.035$ and $l\sim0^\circ.005$ to $\sim20\arcsec$  around $l\sim-0^\circ.050$ and $l\sim-0^\circ.025$.  These filaments are seem to be twisted each other.
The appearance of the MFB is similar to each other in both the emission lines.
M0.014-0.054 is located around the apparent intersection of the MFB and VP.  There is shocked molecular gas abundantly in M0.014-0.054 as shown in Section 4.1. Although the velocity of M0.014-0.054 is similar to that of the VP, it is $V_{\mathrm{LSR}}\sim40$ km s$^{-1}$ different from that of the MFB.  As will be discussed in detail later, M0.014-0.054 is connected to the MFB by the "Bridge" on the PV diagram and a collisionally excited maser spot is also located in M0.014-0.054. These features demonstrate that the MFB and VP collided each other within $10^5$ years and M0.014-0.054 was made in the interaction by the CCC consequently. In addition, M0.014-0.054 is resolved into a series of compact features in these figures.  The strongest feature (labeled A) and second one (labeled B) in the CS emission line maps are centered at $l \sim0.021^\circ, b \sim-0.051^\circ$ (the left circle) and $l \sim0.014^\circ, b \sim-0.052^\circ$  (the right circle), respectively. 

\subsection{Molecular Could Masses of M0.014-0.054, ``Vertical Part",  and ``Molecular Filament Bundle"}
The derivation of molecular cloud mass should suffer from the optical thickness of the observed emission line. The  ratio of the observed brightness temperatures of the C$^{32}$S and C$^{34}$S emission lines is given by 
\begin{equation}
\label{1}
R=\frac{T_\mathrm{B}\mathrm{(C^{32}S)}}{T_\mathrm{B}\mathrm{(C^{34}S)}}
=\frac{f\mathrm{(C^{32}S)}T\mathrm{_{ex}(C^{32}S)}(1-e^{-\tau})}{f\mathrm{(C^{34}S)}T\mathrm{_{ex}(C^{34}S)}(1-e^\frac{-\tau}{22.35})},
\end{equation}
where $T\mathrm{_{ex}}$ and $f$ are the excitation temperature and beam-filling factor of the emission line, respectively. The isotope abundance ratio of $^{32}$S and $^{34}$S in the molecular cloud is assumed to be equal to the natural abundance ratio of $22.35$\footnote{$https://physics.nist.gov/cgi-bin/Compositions/stand\_alone.pl$}.

The mean line intensity ratio in M0.014-0.054 is calculated to be $R\sim9.4$.
Therefore the mean optical thickness of the C$^{32}$S emission line toward M0.014-0.054 is estimated to be $\tau\sim2.1$ assuming that $T\mathrm{_{ex}(C^{32}S)}=T\mathrm{_{ex}(C^{34}S)}$ and  $f(\mathrm{C^{32}S})=f(\mathrm{C^{34}S})$. 
The correction factor for the optical thickness of the C$^{32}$S emission line is  $\frac{\tau}{1-e^{-\tau}}\sim 2.4$. 
The corrected integrated intensity  of the object in  the CS $J=2-1$ emission line is given by
\begin{equation}
\label{2}
I\mathrm{_{corr.}(CS) }=\frac{\tau}{1-e^{-\tau}} \int_{\mathrm{object}}\int T\mathrm{_B (C^{32}S)}dvds.
\end{equation}
The total number of the H$_2$ molecules in the object is given by 
\begin{equation}
\label{3}
\int_{\mathrm{object}}N{\mathrm{_{LTE}(H_2) }}ds=\frac{4.56\times 10^{11} e^{\frac{2.35}{T_\mathrm{ex}}}I\mathrm{_{corr.}(CS)}}
{X\mathrm{(CS)} (1-e^{\frac{-4.7}{T_\mathrm{ex}}})},
\end{equation}
where $N{\mathrm{_{LTE}(H_2) }}$ is the LTE molecular column density.
Here the Einstein $A$ coefficient of the CS $J=2-1$ emission line is assumed to be $A_{21}=2.2\times10^{-5}$ s$^{-1}$.

The excitation temperature of the CS $J=2-1$ emission line is assumed to be $T_{\mathrm{ex}}=80$ K ($T_{\mathrm{K}}=80$ K in \cite{Ao2013}).
The fractional abundance of the CS molecule is assumed to be $X\mathrm{(CS)}=\frac{N\mathrm{(CS)}}{N\mathrm{(H_2)}}=1\times10^{-8}$, which is usually used for molecular clouds in the disk region.  The LTE molecular cloud mass is given by 
\begin{equation}
\label{4}
M_{\mathrm{LTE}}[M_{\odot}]=\mu\int_{\mathrm{object}}N{\mathrm{_{LTE}(H_2) }}ds \simeq2.4\times 10^{-38} T_\mathrm{ex}I\mathrm{_{corr.}(CS)},
\end{equation}
where $\mu$ is  the mean molecular weight per H$_2$ molecule: $\mu = 2.8$ in amu $=2.4\times 10^{-57} M_{\odot}$. The integrated intensity of the whole of M0.014-0.054 in the C$^{32}$S  emission line is  
$\int_{\mathrm{object}}\int_{-40}^{10} T_\mathrm{B}\mathrm{(C^{32}S)}dvds=1.15\times10^{40}\mathrm{[K~ km s^{-1} cm^2]}$. The LTE molecular cloud mass is estimated to be $M_{\mathrm{C32S, LTE}}\simeq5.3\times10^4(\frac{T_\mathrm{ex}}{80})[M_{\odot}]$.

Although the mean optical thickness of the C$^{32}$S emission line is fairly high in M0.014-0.054, that of the C$^{34}$S emission line is estimated to be as small as $\tau\sim\frac{2.1}{22.35}\sim0.09$. We can estimate the molecular cloud mass using the C$^{34}$S emission line data without the  correction for the optical thickness. The integrated intensity of the whole of M0.014-0.054 in the C$^{34}$S emission line is  $\int\int_{-40}^{10} T_\mathrm{B}\mathrm{(C^{34}S)}dvds=1.22\times10^{39}\mathrm{[K~ km s^{-1} cm^2]}$.  The LTE molecular cloud mass is estimated to be $M_{\mathrm{C34S, LTE}}\simeq5.3\times10^4(\frac{T_\mathrm{ex}}{80})[M_{\odot}]$. The LTE molecular cloud mass is consistent with that from the C$^{32}$S  emission line observation. These are summarized in Table 2.

Using the same procedure and the C$^{32}$S emission line data, the LTE molecular cloud masses of the VP and MFB are $M_{\mathrm{C32S, LTE}}\simeq4.3\times10^4(\frac{T_\mathrm{ex}}{80})[M_{\odot}]$ and $M_{\mathrm{C32S, LTE}}\simeq8.4\times10^4(\frac{T_\mathrm{ex}}{80})[M_{\odot}]$, respectively. Meanwhile, using the C$^{34}$S emission line data, the LTE molecular cloud masses of the VP and MFB are $M_{\mathrm{C34S, LTE}}\simeq4.2\times10^4(\frac{T_\mathrm{ex}}{80})[M_{\odot}]$ and $M_{\mathrm{C34S, LTE}}\simeq8.1\times10^4(\frac{T_\mathrm{ex}}{80})[M_{\odot}]$, respectively. These are also summarized in Table 2.

\begin{table}
  \caption{Molecular Could Masses }
  \label{tab:second}
 \begin{center}
    \begin{tabular}{cccccc}
 \hline   \hline
Object&$\bar{\tau}$(C$^{32}$S) &$M_{\mathrm{C32S, LTE}}$&$M_{\mathrm{C34S, LTE}}$&$M_{\mathrm{vir}}$&$\frac{M_{\mathrm{vir}}}{M_{\mathrm{LTE}}}$\\
&&[$M_{\mathrm{\odot}}$]&[$M_{\mathrm{\odot}}$]&[$M_{\mathrm{\odot}}$]&\\
\hline
M0.014-0.054&$2.1$&$5.3\times10^4(\frac{T_\mathrm{ex}}{80})$&$5.3\times10^4(\frac{T_\mathrm{ex}}{80})$&-&-\\
Vertical Part&$1.3$&$4.3\times10^4(\frac{T_\mathrm{ex}}{80})$&$4.2\times10^4(\frac{T_\mathrm{ex}}{80})$&-&- \\
Molecular Filament Bundle&$1.8$&$8.4\times10^4(\frac{T_\mathrm{ex}}{80})$&$8.1\times10^4(\frac{T_\mathrm{ex}}{80})$&-&-\\
\hline
Object A&$3.8$&$-$&$1.3\times10^4(\frac{T_\mathrm{ex}}{80})$&$5.0\times10^3$&0.4\\
Object B&$2.7$&$-$&$1.2\times10^4(\frac{T_\mathrm{ex}}{80})$&-&-\\
 \hline
\end{tabular}
  \end{center}
 Typical error of these masses is estimated to be $\sim30$ \%.
\clearpage
\end{table}

\section{The Cloud-Cloud Collision between the ``Vertical Part" and ``Molecular Filament Bundle"}
As mentioned in the previous section,  the MFB, VP, and M0.014-0.054 are clearly seen in Figures 5a and 5b.  M0.014-0.054 is located positionally in the interacting area between the MFB and VP. 
Figures 6a, 6c, and 6e show the enlarged integrated intensity maps around the interacting area in the C$^{32}$S $J=2-1$,  CH$_3$OH (blended lines around 96.741 GHz), and $^{28}$SiO $J=2-1$  emission lines with the velocity ranges of $V_{\mathrm{LSR}}=10$ to $35$ km s$^{-1}$, respectively.  The velocity ranges include that of the MFB.  The C$^{32}$S data includes the single-dish data.  
On the other hand, Figures 7a, 7c, and 7e show the enlarged integrated intensity maps of the same area in the C$^{32}$S $J=2-1$,  CH$_3$OH, and $^{28}$SiO $J=2-1$  emission lines with the velocity ranges of $V_{\mathrm{LSR}}=-25$ to $0$ km s$^{-1}$, respectively. The velocity ranges include those of the VP and M0.014-0.054. 
As mentioned previously, the SiO and CH$_3$OH emission lines are good tracers of hard ($\Delta V\simeq 30$ km s$^{-1}$) and mild ($\Delta V\simeq10$ km s$^{-1}$) C-shock waves, respectively. 
Although the filaments of the MFB are clearly seen in Figures 6a, they almost disappear in Figures 6c and 6e except for several faint features. The appearances in both shock tracer lines quite resemble each other.  
On the other hand, M0.014-0.054 is prominent in Figures 7a, 7c and 7e. The VP is faint in Figures 7c and 7e although it is identified clearly in Figures 7a.
These indicate that the whole of the MFB and VP did not suffer from any C-shock wave with $\Delta V\gtrsim10$ km s$^{-1}$ but a hard C-shock wave is propagating in M0.014-0.054. 

Figures 6b, 6d and 6f show the PV diagrams along the MFB in the C$^{32}$S $J=2-1$,  CH$_3$OH, and $^{28}$SiO $J=2-1$ emission lines,  respectively. The sampling area is shown as the rectangles in Figures 6a, 6c and 6e. In Figures 6b, the MFB is clearly identified as a long inclined ridge with $V_{\mathrm{LSR}}\sim-20$ at the western end to $30-50$ km s$^{-1}$ at the eastern end.   The MFB almost disappears in the SiO and CH$_3$OH emission lines except for the following several faint features.   In these PV diagrams, M0.014-0.054 is identified around  the angular offset of $-80\arcsec$.   Figures 7b, 7d, and 7f shows the PV diagrams along M0.014-0.054  in the C$^{32}$S $J=2-1$,  CH$_3$OH, and $^{28}$SiO $J=2-1$  emission lines,  respectively.   The sampling area is along M0.014-0.054, which is shown as the rectangles in Figures 7a ,7c, and 7e. 
In these PV diagrams, M0.014-0.054 is clearly identified as a long feature with no clear velocity gradient although the velocity width becomes large around the east end. 
The radial velocity difference between M0.014-0.054 on the VP and the MFB is $\Delta V_{\mathrm{rad}}= \Delta V \cos\phi\sim 40$ km s$^{-1}$, where $\Delta V$ is the collision velocity in 3D space and $\phi$ is the angle between the line-of-sight and  the axis of the collision.  
The shock wave induced by the CCC between the VP and MFB is as strong as bringing the enhancement of the SiO molecule in M0.014-0.054. 

The ``Bridge" features are identified in these PV diagrams of Figures 6 and 7, which are connecting between M0.014-0.054 on the VP and the MFB (arrows, also see a red arrow in Figure 3b). 
The ``Bridge"s also correspond to the faint features seen in Figures 6c and 6e.  They contain shocked molecular gas made by hard C-shock wave.
Recent hydrodynamical simulation studies have shown that such connecting features in PV diagrams are reproduced as a characteristic feature of a CCC  (e.g. \cite{Haworth}, \cite{Haworthb}).  Therefore the ``Bridge" features support the scenario that the MFB and VP  collided each other and made the ``Bridge" with the intermediate velocity between those of the MFB and VP, which should be caused by the momentum exchange of the molecular cloud in them. 
Especially, the feature around the angular offset of $-60\arcsec$ is more prominent than that around the angular offset of $-80\arcsec$ around in these emission lines.  SiO molecules have been  known to be taken in the mantle of interstellar dusts within $10^5$ years and disappear from interstellar gas  as mentioned previously. 
If the CCC was occurs once in such past, the difference between the``Bridge"s may be made by the elapsed time from the collision.  

However, if the  MFB and VP collided with the proper motion velocity of $V=40\times \tan\phi$ km/s before $10^5$ yr, one molecular cloud is now separating from another molecular cloud by $\Delta d\sim4.1\times\tan\phi$ pc.
The separation is corresponding to the angular separation of $\Delta\theta\sim0.03^\circ\times\tan\phi$ at the distance of 8 kpc, which can be easily distinguished in this observation although $\phi$ is unknown. However, such large angular separations are not observed in the ``Bridge"s connecting two clouds. 
There are at least two possibilities to explain it.
As will be mentioned in Section 7, mGauss magnetic fields run along these filaments. In this case, the molecular cloud can move along the filaments but cannot move crossing them. Therefore, the molecular cloud in the Bridge would be fixed around the collision area by the magnetic field which are probably tangled by the collision.
On the other hand, $10^5$ yr is the upper limit of the SiO molecules surviving in the interstellar space before re-absorption into dust and not the elapsed time from the CCC. 
Therefore, this may suggest the possibility that the HMCs are formed in a shorter time than the surviving upper limit of the SiO molecule although it is required that the angle $\phi$ is somewhat small.
\begin{figure}
\begin{center}
\includegraphics[width=18cm,  bb=0 0 702.63 403.19]{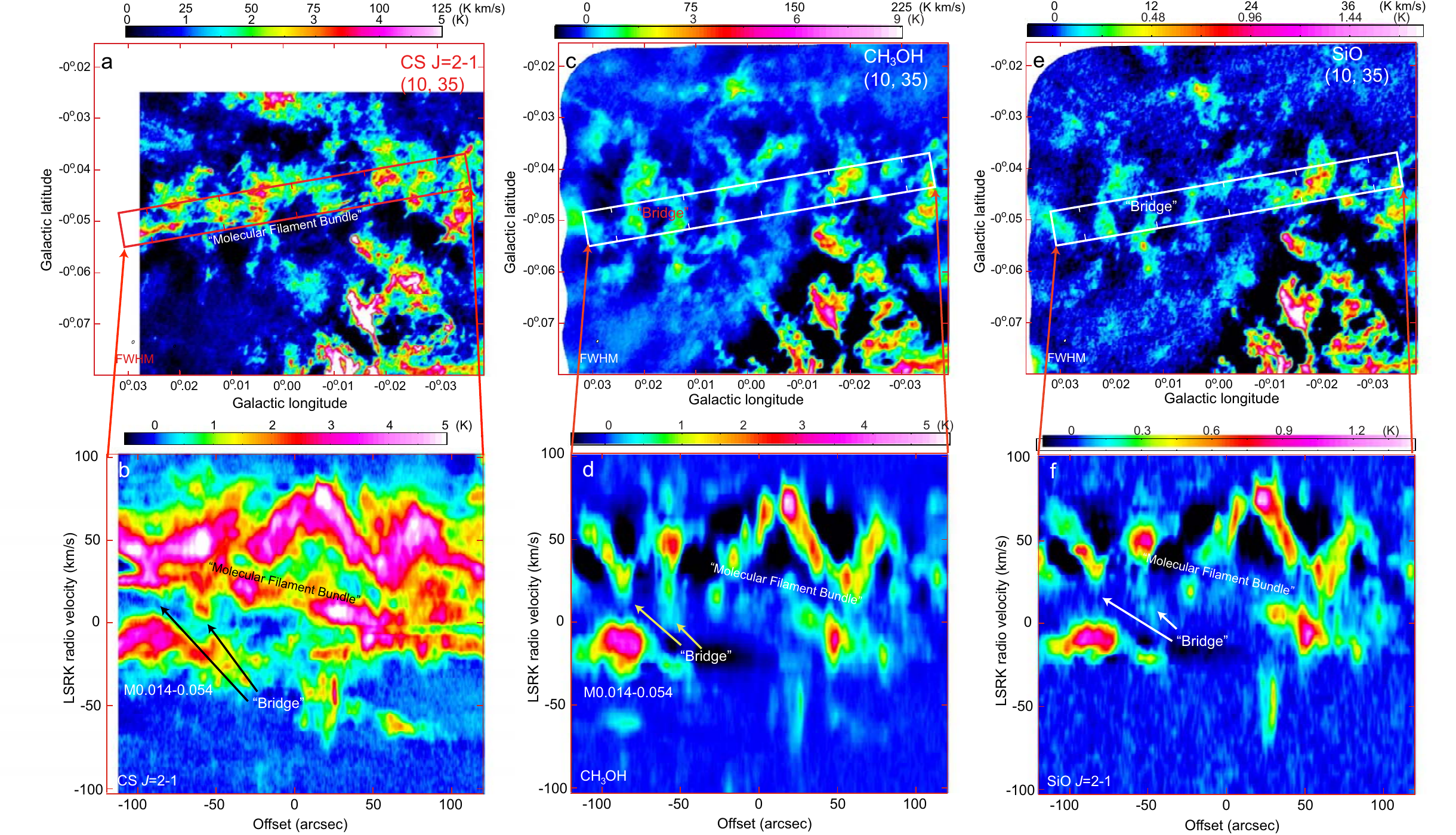}
 \end{center}
 \caption{{\bf a} Enlarged integrated intensity map around the "Molecular Filament Bundle"  in the C$^{32}$S $J=2-1$ emission line with the velocity ranges of $V_{\mathrm{LSR}}= 10$ to $35$ km s$^{-1}$. 
 {\bf b} PV diagram along the  "Molecular Filament Bundle" in the C$^{32}$S $J=2-1$ emission line. The sampling area is shown as the rectangles in {\bf a}. The C$^{32}$S maps are after the combining with the single-dish data.
 {\bf c}  Enlarged integrated intensity map in the CH$_3$OH emission of $V_{\mathrm{LSR}}=10$ to $35$ km s$^{-1}$. 
 {\bf d} PV diagram along the "Molecular Filament Bundle" in the CH$_3$OH emission line. The sampling area is shown as the rectangles in {\bf c}. 
There is a faint component with $\Delta V \sim -50$  km s$^{-1}$ from M0.014-0.054. This is the contamination of the CH$_3$OH$ J_{K_a, K_c}=2_{1,1}-1_{1,0}$E emission line at $96.755507$ GHz.
 {\bf e}  Enlarged integrated intensity map in the $^{28}$SiO $J=2-1$ emission of $V_{\mathrm{LSR}}=10$ to $35$ km s$^{-1}$. 
  {\bf f} PV diagram along along the "Molecular Filament Bundle"  in the  $^{28}$SiO $J=2-1$ emission line. The sampling area is shown as the rectangles in {\bf e}.}
 \label{Fig6}
 \clearpage
\end{figure}

\begin{figure}
\begin{center}
\includegraphics[width=18cm, bb=0 0  688.35 403.31]{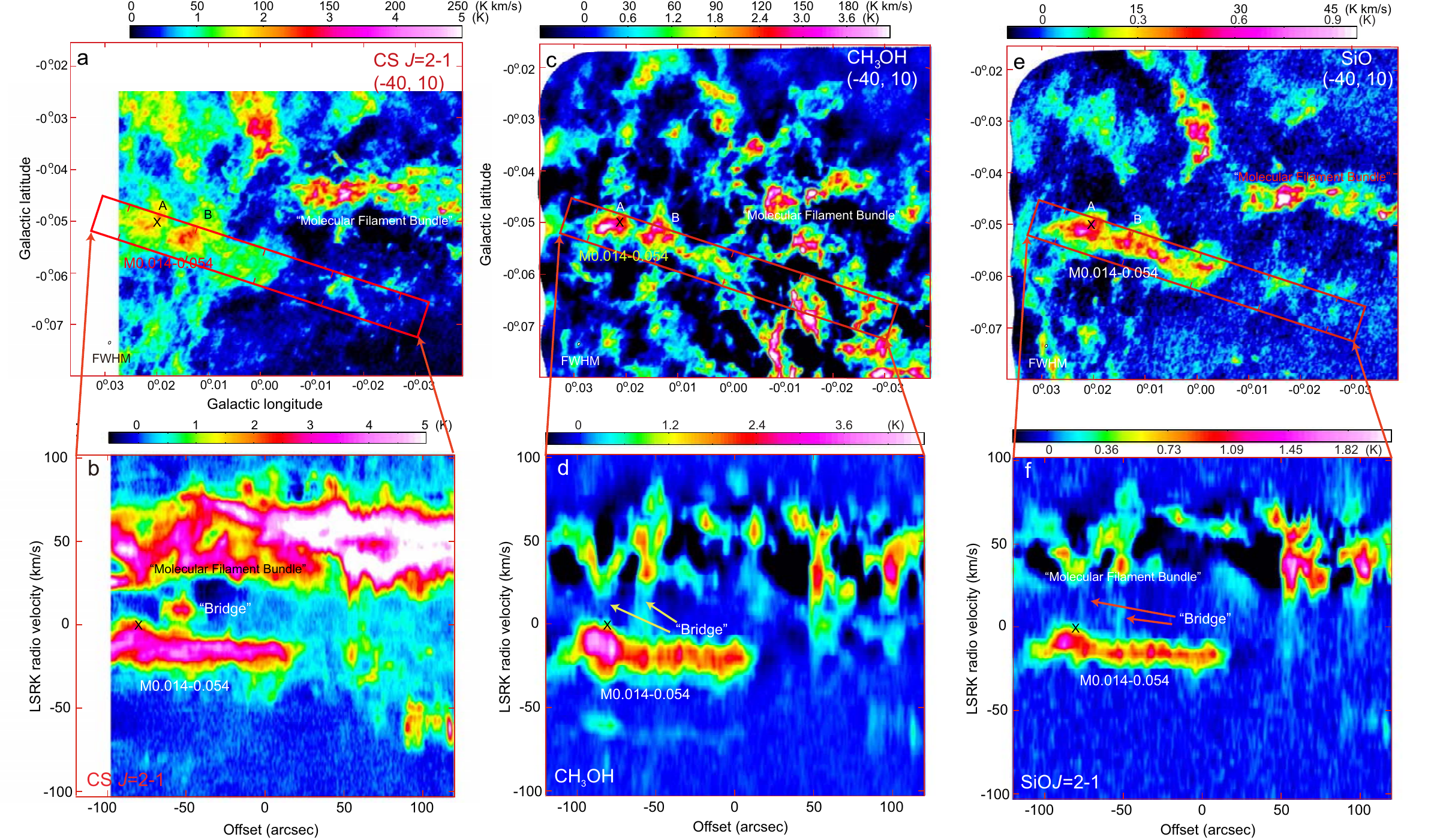}
 \end{center}
 \caption{
 {\bf a} Enlarged integrated intensity map around M0.014-0.054  in the C$^{32}$S $J=2-1$ emission line with the velocity ranges of ${\bf a}~V_{\mathrm{LSR}}=-40$ to $10$ km s$^{-1}$. The cross shows the position of the collisionally excited CH$_3$OH maser emission at 36.2 GHz in the same velocity range \citep{Yusef-Zadeh2013}.
 {\bf b} PV diagram along M0.014-0.054  in the C$^{32}$S $J=2-1$ emission line. 
 The sampling area is shown as the rectangles in {\bf a}. The C$^{32}$S maps are after the combining with the single-dish data. The cross shows the position and velocity of the collisionally excited CH$_3$OH maser emission at 36.2 GHz \citep{Yusef-Zadeh2013}.
 {\bf c}  Enlarged integrated intensity map in the CH$_3$OH emission of ${\bf a}~V_{\mathrm{LSR}}=-40$ to $10$ km s$^{-1}$. 
 {\bf d} PV diagram along M0.014-0.054  in the  CH$_3$OH emission line. The sampling area is shown as the rectangles in {\bf c}. There is a faint component with $\Delta V \sim -50$  km s$^{-1}$ from M0.014-0.054. This is the contamination of the CH$_3$OH$ J_{K_a, K_c}=2_{1,1}-1_{1,0}$E emission line at $96.755507$ GHz.
   {\bf e}  Enlarged integrated intensity map in the $^{28}$SiO $J=2-1$ emission of ${\bf a}~V_{\mathrm{LSR}}=-40$ to $10$ km s$^{-1}$. 
  {\bf f} PV diagram along M0.014-0.054  in the  $^{28}$SiO $J=2-1$ emission line. The sampling area is shown as the rectangles in {\bf e}.
}
 \label{Fig7}
  \clearpage
\end{figure}

The crosses in Figures 7a, 7c, and 7e show the position, $l=0^\circ.02055~b=-0^\circ.05013$, of the CH$_3$OH $J=4_{-1}-3_0$ maser emission spot at 36.2 GHz in the velocity range of $V=0.0\pm8.3$ km s$^{-1}$ (\cite{Yusef-Zadeh2013}). The maser emission spot is located in the vicinity of the strongest intensity peak of the maps.  While the crosses in Figures 7b, 7d, and 7f show the position and velocity of the CH$_3$OH $J=4_{-1}-3_0$ maser emission spot in the PV diagrams. The velocity range of the maser emission spot overlaps with the positive velocity edge of M0.014-0.054.
Because the CH$_3$OH maser emission is known to be excited collisionally by shock wave propagating in molecular cloud, these facts indicate that the CCC  between the MFB and VP is on-going. 

Figures 6b and 7b also shows the large feature with $V_{\mathrm{LSR}}\sim20$ to $80$ km s$^{-1}$, which  has been mentioned  in Section 4.2. This feature seems to be made of several ridge-like features. Although the ridge-like features are also identified in Figures 6d, 6f, 7d, and 7f, the spines of the ridges seem to be enhanced.
In addition, a faint feature is also identified around the angular extent of M0.014-0.054 and velocity range of $V_{\mathrm{LSR}}=-50$ to$-70$ km s$^{-1}$ in the PV diagrams of the CH$_3$OH emission line (see Figures 6d and 7d). The feature is the contamination by the CH$_3$OH  $J_{K_a, K_c}=2_{1,1}-1_{1,0}E$ emission line ($96.755507$ GHz) because this feature is not detected  in the C$^{32}$S $J=2-1$ emission line (see Figures 6b and 7b).
  
\section{``Hot Molecular Core"s in the Interacting Area}
\subsection{Identification of Hot Molecular Cores in M0.014-0.054}
Figure 8a shows the continuum map  at 86 GHz of M0.014-0.054. The noise level of the map is $1\sigma=0.01$ K in $T_\mathrm{B}$.
M0.014-0.054  is detected as a series of the compact objects, A, B, C, D, and E. The eastern two objects are prominent, which correspond to ``A" and ``B" shown in Figures 5a and 5b. The cross in the map indicates the position, $l=0^\circ.02055~b=-0^\circ.05013$, of the CH$_3$OH $J=4_{-1}-3_0$ maser emission spot at 36.2 GHz in the velocity range of $V=0.0\pm8.3$ km s$^{-1}$ (\cite{Yusef-Zadeh2013}).  This is located in the object A. 
The mean brightness temperatures of the continuum emission in the objects A and B are $\bar{T}_\mathrm{B, cont}\sim0.030\pm0.002$ K and $\bar{T}_\mathrm{B, cont}\sim0.020\pm0.001$ K, respectively. The integration areas are shown as the circles in the map.
The brightness temperature ratio of the recombination line to continuum emission is given by 
the formula, $T_\mathrm{B, line}/T_\mathrm{B, cont}\sim3\times10^4[T_\mathrm{e}/\mathrm{K}]^{-1.65}[\nu/\mathrm{GHz}]^{1.1}$ in the assumption of LTE and optically thin condition (e.g. \cite{Mezger}). Therefore, the ratio at 86 GHz of ionized gas with up to $T_\mathrm{e}\lesssim1\times10^4$ K is estimated to be larger than unity, $T\mathrm{_B (H42\alpha)}/T_\mathrm{B, cont}>1$. 
 The mean brightness temperatures of the H42$\alpha$ recombination line are expected to be $\bar{T}\mathrm{_B (H42\alpha)}> 0.030$ K in the object A and $\bar{T}\mathrm{_B (H42\alpha)}> 0.020$ K in the object B, respectively.
Figure 8b shows the integrated intensity map with the velocity range of $-25$ to $0$ km s$^{-1}$ 
of the H42$\alpha$ recombination line. The velocity width is as large as that of the recombination line from the ionized gas at $T_\mathrm{e}\sim 1\times10^4$ K.
The upper limits of the mean brightness temperatures in the H42$\alpha$ recombination line are $5\sigma = 0.019$ K at the object A and $5\sigma = 0.019$ K at the object B, respectively. The integration areas are the same as those mentioned above. The objects A and B are not detected in the map.
There is no ionized gas in the objects A and B of M0.014-0.054 above the detection limit of our observation. 
In the cases of the objects C, D, and E, we cannot exclude the existence of ionized gas by this procedure because of their weak intensities. 
However, the low $T_\mathrm{B}$[C$_2$H]/$T_\mathrm{B}$[c-C$_3$H$_2$] ratios suggest that the objects C, D, and E have no ionized gas as will be discussed in Subsection 6.4.

There are two possibilities to explain the situation that the ionized gas does not exist although the mm-wave continuum is detected. 
One possibility is that these continuum emissions are made by the low frequency extension of the dust emission shown in the JCMT map of Figure 2l. 
As mentioned previously,  the continuum emissions  of the objects A and B at 86 GHz correspond to ``Double peaks" of the $850~\mu$m dust emission with $\bar{T}_\mathrm{B}\sim0.3$ K (see Figure 2l).  
If the dust $\beta$ is $\sim2$, the mean brightness temperatures  at 86 GHz of the objects A and B are consistent with the dust emission at the low frequency. In addition, the objects C, D, and E also have the corresponding components in the $850~\mu$m dust emission (see Figure 2l). 
Another possibility is that this emission is an artifact by contamination from other molecular emission lines. If so, the appearance of the 86 GHz continuum emission in Figure 8a should resemble those in the other molecular emission lines. However, they do not always resemble that of the 86 GHz continuum emission. Therefore, the second possibility seems unlikely. These will be discussed in the followings. 

\begin{figure}
\begin{center}
\includegraphics[width=17cm,  bb=0 0  500.5 505.3]{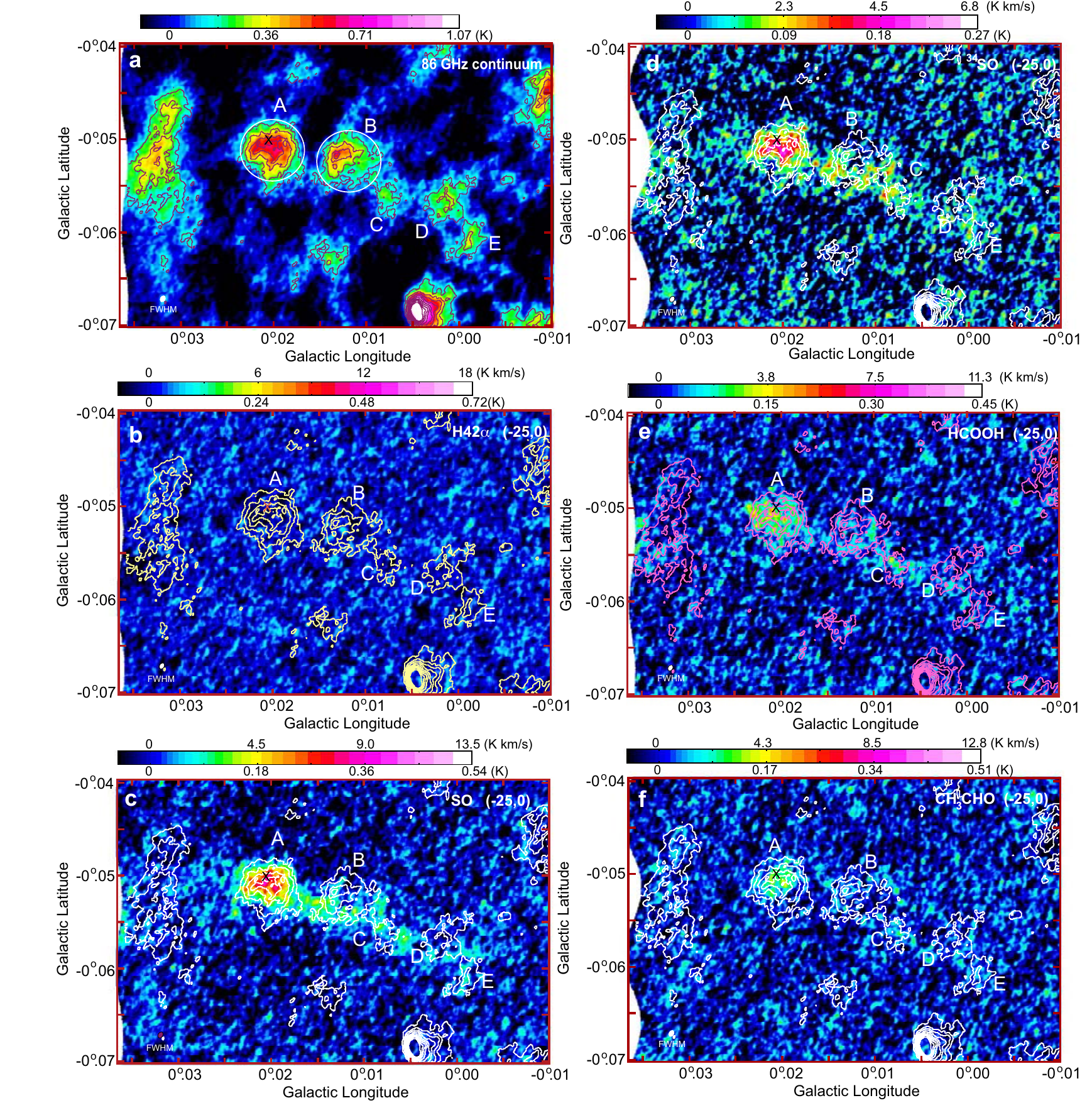}
 \end{center}
 \caption{
 {\bf a} Continuum map of M0.014-0.054 at 86 GHz. The first contour level and contour interval are both 0.014 K. The cross indicates the position, $l=0^\circ.02055~b=-0^\circ.05013$, of the CH$_3$OH $J=4_{-1}-3_0$ maser emission spot at 36.2 GHz in the velocity range of $V=0.0\pm8.3$ km s$^{-1}$ (\cite{Yusef-Zadeh2013}).  
{\bf b} Integrated intensity map with the velocity range of $-25$ to $0$ km s$^{-1}$ ($1\sigma=0.029$ K) of the H42$\alpha$ recombination line. The continuum map at 86 GHz (contours) is superimposed on it.  {\bf c} Integrated intensity map of M0.014-0.054 in the $^{32}$SO~$N,J=2,2-1,1$ emission line with the velocity ranges of ${\bf a}~V_{\mathrm{LSR}}=-25$ to $0$ km s$^{-1}$.   {\bf d} Integrated intensity map in the $^{34}$SO~$N,J=2,3-1,2$ emission line with the velocity ranges of ${\bf a}~V_{\mathrm{LSR}}=-25$ to $0$ km s$^{-1}$. {\bf e} Integrated intensity map in the HCOOH~$J_{K_a, K_c}=4_{1,4}-3_{1,3}$ emission line. {\bf f} Integrated intensity map in the CH$_3$CHO~$J_{K_a, K_c}=5_{2,3}-4_{2,2}E$ emission line.}
 \label{Fig8}
  \clearpage 
\end{figure}

Figures 8c and 8d  show the enlarged integrated intensity maps of M0.014-0.054 in the SO  and $^{34}$SO emission lines with the velocity ranges of $V_{\mathrm{LSR}}=-25$ to $0$ km s$^{-1}$, respectively. As mentioned previously, the SO and $^{34}$SO emission lines are good tracers of the HMC. A compact feature in the SO and $^{34}$SO emission lines is located in the object A.  In contrast, these emissions are associated with the object B but not centered in it. Or rather, these emissions traces the southern limb of the object. These suggest that the object B has a hollow structure in these molecules.  The objects A and B would be HMCs despite some different structures.
The appearances in these emission lines resemble those in the HC$^{15}$N emission line which is shown in Figure 9e. 
Chemical models of HMCs
suggest that the HC$^{15}$N emission line is enhanced in HMCs with the age of $\gtrsim10^5$ years (e.g. see Figure 11 in \cite{Stephan}). Therefore star formation activity may have fairly advanced in the HMCs.
In addition, faint features in the SO, $^{34}$SO and HC$^{15}$N emission lines are also detected around the objects C, D, and E. 
This may suggest that these objects are in the similar evolutional stage to the objects A and B.

Figures 8e and 8f  show the enlarged integrated intensity maps of M0.014-0.054 in the HCOOH and CH$_3$CHO(96.475523GHz) emission lines with the velocity ranges of $V_{\mathrm{LSR}}=-25$ to $0$ km s$^{-1}$, respectively.
Although the emissions of the HCOOH and CH$_3$CHO emission lines are very weak, they are also centered in the object A.  The association with the other objects is marginal.
Figure 10c shows the wide field peak intensity map of the whole area in the HCOOH emission line.  The emissions of HCOOH are identified only on several spots in the 50MC except for in M0.014-0.054. These spots seem to correspond to HMCs identified by Miyawaki et al (2020). The concentration to HMCs of the HCOOH emission should be remarkable. The HCOOH emission line may be a good tracer of HMCs.
Although the distribution in the CH$_3$CHO emission line is not clear in the 50MC because of the contamination of the C$^{34}$S emission line, the similar concentration might be seen.

\begin{figure}
\begin{center}
\includegraphics[width=17cm,   bb=0 0  498.97 503.98]{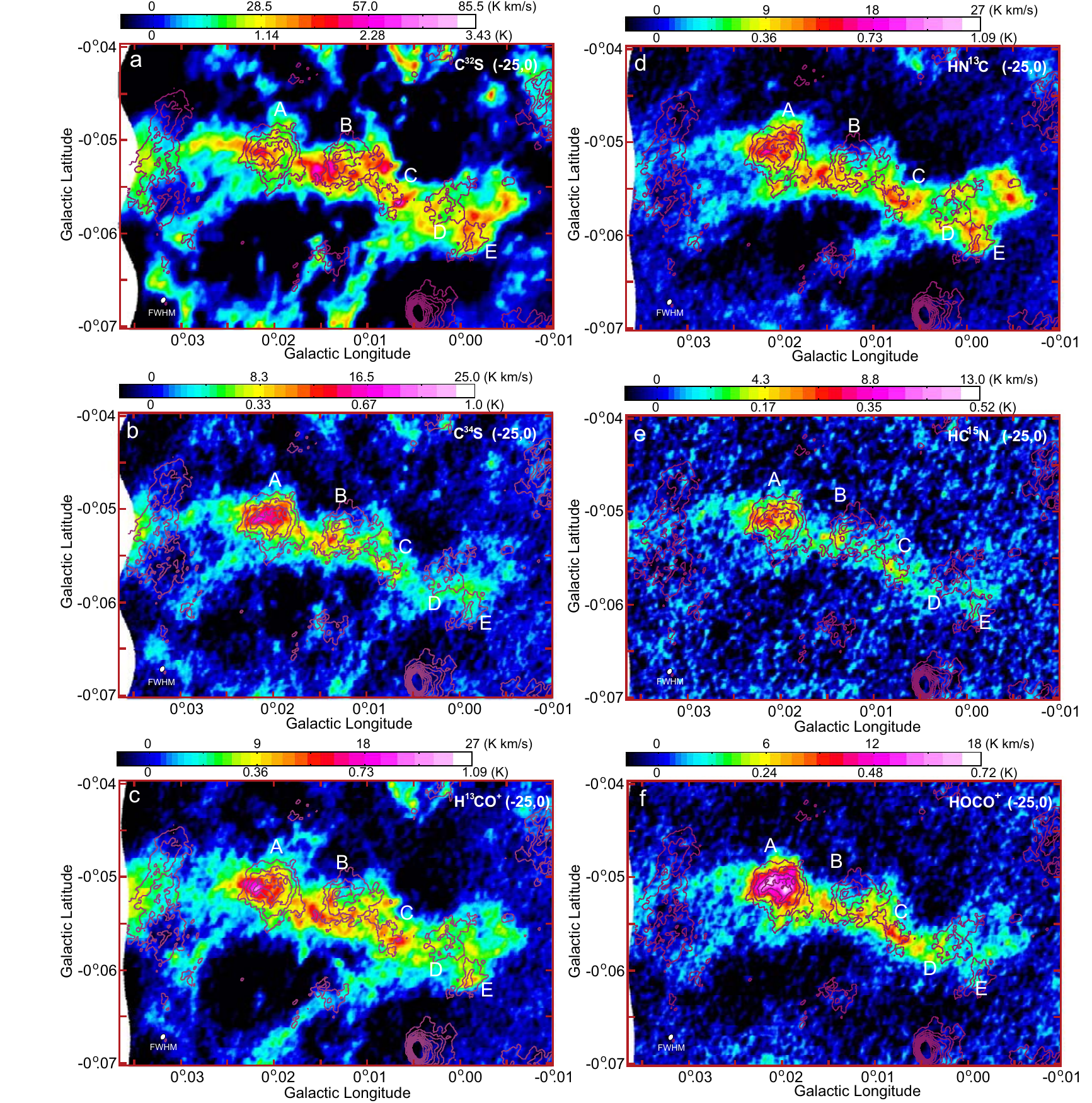}
 \end{center}
 \caption{ Enlarged integrated intensity map of M0.014-0.054 of the C$^{32}$S $J=2-1$ emission line with the velocity ranges of ${\bf a}~V_{\mathrm{LSR}}=-25$ to $0$ km s$^{-1}$. Contours show the continuum map at 86 GHz.  The first contour level and contour interval are both 0.014 K.  {\bf b} Map of the C$^{34}$S $J=2-1$ emission line. {\bf c} Map of the H$^{13}$CO$^+$ $J=1-0$ emission line. {\bf d} Enlarged integrated intensity map of the HN$^{13}$C~$J=1-0~F=1-1$ emission line. 
 {\bf e} Map of the HC$^{15}$N~$J=1-0$ emission line. {\bf f} Map of the HOCO$^+$~$J_{K_a, K_c}=4_{0,4}-3_{0,3}$ emission line.  }
 \label{Fig9}
  \clearpage
\end{figure}
\begin{figure}
\begin{center}
\setcounter{figure}{8}
\includegraphics[width=17cm,   bb=0 0 498.19 336.06]{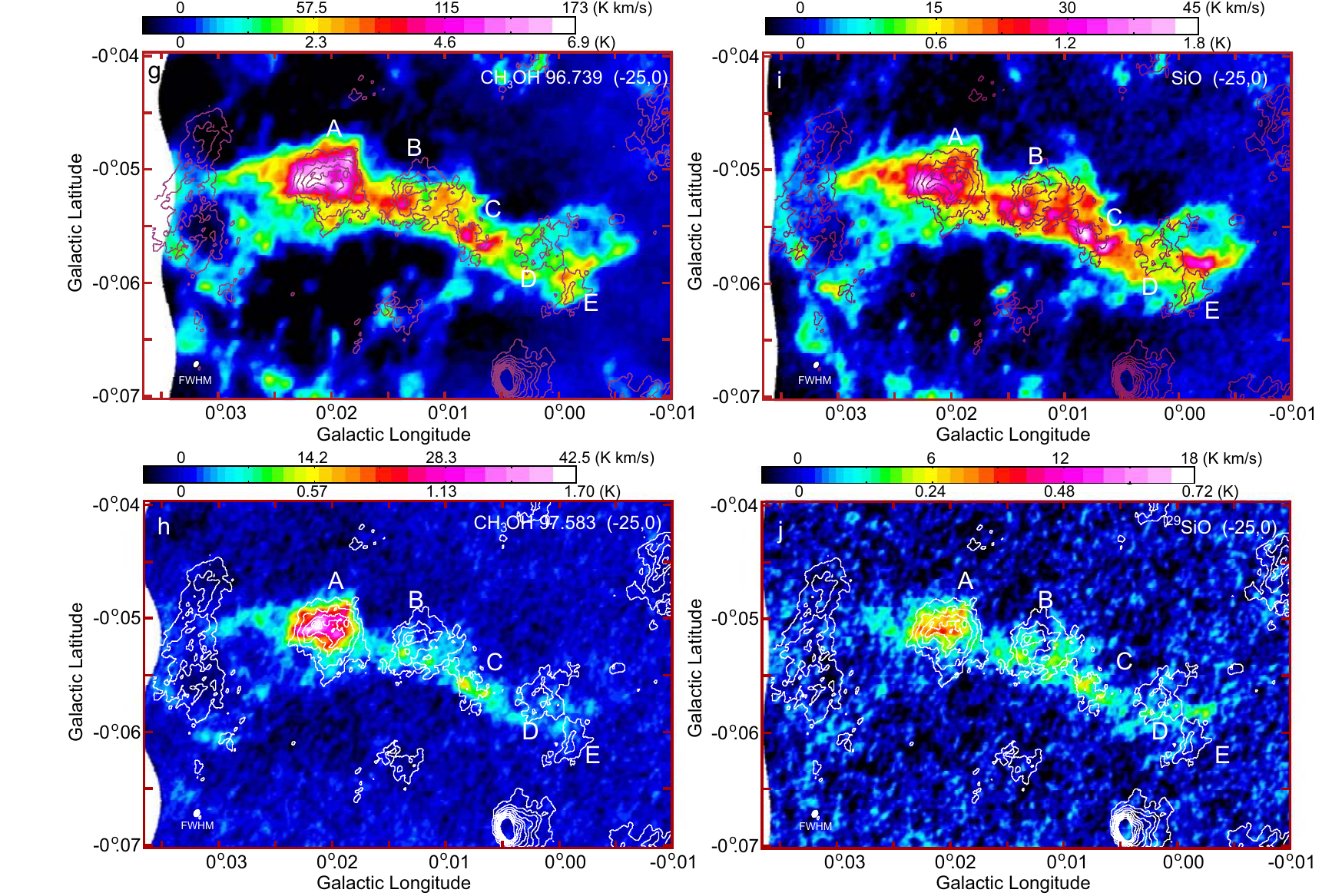}
 \end{center}
 \caption{(Continued.)
 {\bf g} Map of the CH$_3$OH blended emission lines around 96.741 GHz. Contours show the continuum map at 86 GHz.  The first contour level and contour interval are both 0.014 K. {\bf h} Map of the CH$_3$OH emission lines at 96.583 GHz.  {\bf i} Map of the $^{28}$SiO~$J=2-1$ emission line. {\bf j} Map of the $^{29}$SiO~$J=2-1$ emission line.   }
 \label{Fig9b}
  \clearpage
\end{figure}
 
\subsection{Physical Properties of Hot Molecular Cores}
Figures 9a, 9b, 9c, and 9d show the enlarged integrated intensity maps  of M0.014-0.054  in the C$^{32}$S, C$^{34}$S, H$^{13}$CO$^+$, and  HN$^{13}$C emission lines with the velocity ranges of $V_{\mathrm{LSR}}=-25$ to $0$ km s$^{-1}$.  The image in the C$^{32}$S  emission line has a ridge-like feature along the distribution of the objects A-E in the continuum map, which does not resemble the images in the other emission lines as mentioned previously. The image would suffer from  large optical thickness of the C$^{32}$S emission line.  The images in the C$^{34}$S, H$^{13}$CO$^+$ and HN$^{13}$C emission lines resemble each other. These emissions have  strong intensity peaks centered at the object A. Meanwhile, the emissions trace the southern limb of the object B rather than the continuum emission peak. There are also their emissions associated with the objects C, D, and E. 
These similar images would be caused by tracing the similar number density of Hydrogen molecules, $n({\mathrm H}_2)\sim10^5$ cm$^{-3}$. 
Figure 9e shows the map in the HC$^{15}$N emission line with the same velocity range. The critical density of this emission line, $n({\mathrm H}_2)_{\mathrm{crit}}\sim10^7$ cm$^{-3}$, is higher than those of the others. This emission becomes strong at the object A, indicating that the object is filled by a very dense molecular cloud. While, although there is the emission along the south limb of the object B, it dose not centered at the object. The center part of the object B is not filled by such very dense molecular cloud. 

The LTE molecular cloud masses of the objects A and B are estimated from the C$^{32}$S and C$^{34}$S emission line observations using the procedure shown in Section 4.4. 
The mean intensity ratios are calculated to be $R\sim6.3$ in the object A and $R\sim8.1$ in the object B, respectively. Then the optical thicknesses of the C$^{32}$S emission line are $\tau\sim3.8$ in the object A and $\tau\sim2.7$ in the object B, respectively. While those of the C$^{34}$S emission line are $\tau\sim3.8/22.35=0.17$ in the object A and $\tau\sim2.7/22.35=0.12$ in the object B, respectively. Because the C$^{32}$S emission line of the object is fairly thick, the LTE molecular cloud masses of the objects A and B are estimated from  the C$^{34}$S emission line data using the optical thickness correction factors of $\frac{\tau}{1-e^{-\tau}}\sim1.09$ and $\sim1.06$, respectively. The integration velocity range is $V_{\mathrm{LSR}}=-40$ to $10$ km s$^{-1}$.
The LTE molecular cloud masses of the objects A and B are $M_\mathrm{LTE} =1.3\times10^4 (\frac{T_\mathrm{ex}}{80}) M_\odot$ and $M_\mathrm{LTE} =1.2\times10^4 (\frac{T_\mathrm{ex}}{80}) M_\odot$, respectively. 
Although these masses are consistent with the LTE masses of the HMCs detected in the 50MC,  $\bar{M}_\mathrm{LTE} \sim8\times10^3(\frac{T_\mathrm{ex}}{80}) M_\odot$ (\cite{Miyawaki}), they are much larger than those of the molecular cloud cores detected in the 50MC, $\bar{M}_\mathrm{LTE} \sim2\times10^2 (\frac{T_\mathrm{ex}}{50}) M_\odot$ (\cite{Uehara}).

The mean physical radius of the object A is estimated to be $R_\mathrm{HWHM}= 0.33$ pc as the half width at half maximum (HWHM) by 2-dimensional Gaussian fit for the C$^{34}$S image. The FWHM velocity width is estimated to be  $V_\mathrm{FWHM} = 8.5\pm0.7 $ km s$^{-1}$ by 1-dimensional Gaussian fit for the  the C$^{34}$S line profile. Using the following formula for a spherical object, 
\begin{equation}
\label{2}
M_{\mathrm{vir}}[M_{\odot}]=210\times \Delta V_{\mathrm{FWHM}}[\mathrm{km s}^{-1}]^2 R[\mathrm{pc}]
\end{equation} 
the virial mass of the object A is estimated to be $M_\mathrm{vir}=5.0\times10^3 M_\odot$.  
It is difficult to estimate the virial mass of the object B using this formula since the shape of the C$^{34}$S image is not spherical (see Figure 9b).  Because the CCC is ongoing around the object A as mentioned previously, the observed velocity width may be widened, and the derived viral mass should be on the upper limit.
Therefore, the virial parameter of the object A is estimated to be $M_\mathrm{vir}/M_\mathrm{LTE}\lesssim1$.  Although the fractional abundance of the CS molecule and the excitation temperature of the CS emission line have large ambiguity, the object A could be bound gravitationally.  These masses are also summarized in Table 2.

The flux density at 850 $\mu$m in the object A is $S_\nu\sim240$ Jy (see Figure 2l). Because the beam size of JCMT is not sufficiently small to resolve the structure seen in the map and there are other components in the line-of-sight (see Figures 3 and 4), the flux density should be the upper limit.  Assuming that the dust temperature is 33 K (see Fig. 3 in \cite{Etxaluze}) and the gas-to-dust ratio is 100, the gas mass is estimated to be $M\sim7\times10^3 M_\odot$.  This mass is consistent with the LTE molecular cloud mass and larger than the virial mass. 
This also indicates that the object A is bound gravitationally. 
\begin{figure}
\begin{center}
\includegraphics[width=14cm,  bb=0 0 466.81 655.24]{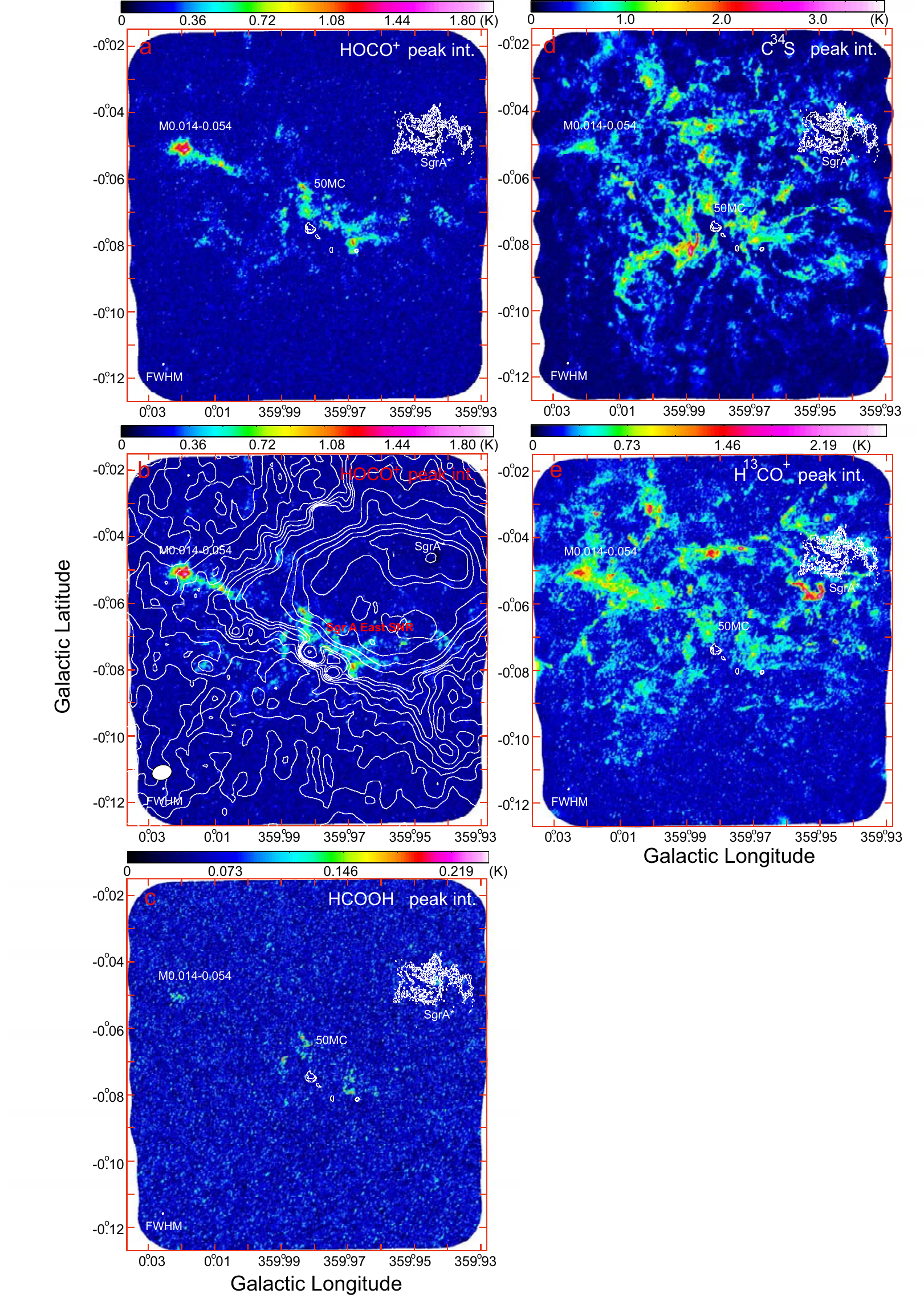}
 \end{center}
 \caption{{\bf a} and {\bf b} Peak intensity maps of the field including the 50MC and Sgr A in the HOCO$^+$ emission line. {\bf c},  {\bf d}, and {\bf e} Peak intensity maps in the  HCOOH, C$^{34}$S, and H$^{13}$CO$^+$ emission lines, respectively.
The velocity range of the maps is $V_{\mathrm{LSR}}=-150$ to $150$ km s$^{-1}$. The FWHM beam sizes are shown in the lower left corners. 
The contours in {\bf a, c, d} and {\bf e} show the continuum emission at 86 GHz for comparison. The contour levels are $(1, 2, 8, \mathrm{and}~32)\times 3.5$ K. The contours in {\bf b} show the continuum emission at 1.44 GHz  for comparison (\cite{Yusef-Zadeh1984}).  The contour levels are $(1, 2, 3, 4, 5, 6, 7, 8, 9, 10, 15, 20, 30, 40, \mathrm{and}~50)\times 24$ K.}
 \label{Fig10}
  \clearpage
\end{figure}

\subsection{Chemical Properties of Hot Molecular Cores}
 \subsubsection{HOCO$^+$ Ion}
The HOCO$^+$ emission line has been detected toward the molecular clouds in the CMZ including the 50MC (e.g. \cite{Minh}).
Figure 9f shows the map of M0.014-0.054 in the HOCO$^+$ emission line with the velocity range of $V_{\mathrm{LSR}}=-25$ to $0$ km s$^{-1}$. This emission is clearly centered on the object A. The distribution closely resembles those of the C$^{34}$S and H$^{13}$CO$^+$ emission lines (see Figures 9b and 9c). The C$^{34}$S emission line is considered to be a faithful tracer of the dense molecular cloud with $n({\mathrm H}_2)_{\mathrm{crit}}\sim10^5$ cm$^{-3}$. The effective critical density of the H$^{13}$CO$^+$ emission line is similar to that of the C$^{34}$S emission line. Although this similarity is probably why the distribution of the C$^{34}$S and H$^{13}$CO$^+$ emission lines resemble each other,
it cannot be explained why that of the HOCO$^+$ emission line resembles them. 

Figure 10a shows the wide field peak intensity map of the whole area of this observation, which includes the 50MC and Sgr A$^\ast$, in the HOCO$^+$ emission line.  The velocity range is $V_{\mathrm{LSR}}=-150$ to $150$ km s$^{-1}$. The emission in M0.014-0.054, especially the object A, is the most prominent in the velocity range and area although some emissions are identified in the 50MC. The distribution might originate with the spatial variation of the fractional abundance of the HOCO$^+$ ion.  
Figures 10d and 10e show the wide field peak intensity maps in the C$^{34}$S and  H$^{13}$CO$^+$ emission lines, respectively.
In the C$^{34}$S emission line,  the 50MC is the most prominent molecular cloud in the area, although M0.014-0.054 is relatively inconspicuous (also see Figure 5 in \cite{Uehara}). On the other hand, the distribution of the H$^{13}$CO$^+$ emission line in M0.014-0.054  is more conspicuous than that in the 50MC.  Although the effective critical densities of these emission lines are similar, the distributions have distinctive difference. The distribution of the HOCO$^+$ emission line  relatively resembles that of the H$^{13}$CO$^+$ emission line.

There are some possibilities of the formation of the HOCO$^+$ ion which could make the distributions mentioned above. 
An ion-molecule reaction; $$\mathrm{HCO}^++\mathrm{OH}\rightarrow \mathrm{HOCO}^+ +   \mathrm{H},$$ might be possible (e.g. \cite{Fontani}) because OH radical should be abundant  and the gas kinetic temperature is high in the Sgr A region.  Because CO molecule is also abundant in the region, the reaction consuming the HOCO$^+$ ion, 
$$\mathrm{HOCO}^++\mathrm{CO} \rightarrow \mathrm{HCO}^++    \mathrm{CO}_2,$$ 
is also expected in the region (e.g. \cite{Sakai}). 
In this case, the distribution of the HOCO$^+$ emission line would  resemble that  of the H$^{13}$CO$^+$ emission line.  

The HOCO$^+$ ion may be formed by the protonated reaction of CO$_2$ molecule.  H$_3^+$ ion is abundant by the abundant cosmic ray in the Sgr A region (e.g. \cite{OkaT}).  CO$_2$ molecules are evaporated extensively from the dust mantle by high gas kinetic temperature in M0.014-0.054 ($T_\mathrm{K}\simeq80$ K in \cite{Ao2013}). Therefore an ion-molecule reaction forming the HOCO$^+$ ion, 
$$\mathrm{H}_3^++   \mathrm{CO}_2\rightarrow \mathrm{HOCO}^++ \mathrm{H}_2,$$
is also expected in the region (e.g. \cite{Sakai}).   In this case, it is not necessary that the distribution of the HOCO$^+$ emission line resemble that of the H$^{13}$CO$^+$ emission line. Note that these reactions  are not mutually exclusive and other reactions forming and destroying HOCO$^+$ ion are also possible.

Figure 10b shows the comparison between the peak intensity map of the HOCO$^+$ emission line and 1.44 GHz continuum emission (\cite{Yusef-Zadeh1984}). The features in the 50MC observed in the HOCO$^+$ emission line are located only along the southeast edge of Sgr A East SNR. This suggests that the physical interaction between the 50MC and Sgr A East SNR forms the HOCO$^+$ ions or at least assists the formation. There are two possibilities explaining the scenario. The first one may be concerned in shock chemistry by the SNR. However, it is hard to specify the mechanism of the formation immediately.  In the second one, abundant OH radicals made by cosmic ray of the SNR may form the ions through the reaction mentioned above (also see \cite{Tielens}). 

On the other hands, the formation scenario of the features observed in M0.014-0.054 is not concerned probably in  SNRs because there is no known SNR around it. However, the low frequency continuum emission associated with the south edge of M0.014-0.054 is seen. This is identified more clearly at 330 MHz as mentioned in Sec.2 (see Figure 1d). A possibility explaining the continuum emission would be that the CCC between the VP and MFB accelerates cosmic ray. The strong magnetic field (see Section 7) and large collision velocity (see Section 5) may enable such acceleration around M0.014-0.054 although it has not been observed in the disk region. The accelerated cosmic ray would increase the abundance of OH radicals simultaneously (also see \cite{Tielens}). Therefore the HOCO$^+$ ions would be formed through second scenario mentioned above. 

 \subsubsection{CH$_3$OH and SiO Molecules}
As mentioned previously, the C-type shock wave propagating in the molecular cloud enhances the abundances of the CH$_3$OH and SiO molecules. Figures 9g and 9h  show the enlarged integrated intensity maps of M0.014-0.054 in the CH$_3$OH (blended lines around 96.741 GHz) and CH$_3$OH (97.583 GHz) emission lines with the velocity ranges of $V_{\mathrm{LSR}}=-25$ to $0$ km s$^{-1}$, respectively. However, the higher $J$ transition lines of CH$_3$OH molecule, $J_{K_a, K_c}=6_{-2,5}-7_{-1,7}$ E at 85.568 GHz , $J_{K_a, K_c}=7_{2,6}-6_{3,3}$ A$_{--}$ at 86.616 GHz,  $J_{K_a, K_c}=21_{6,16}-22_{5,17}$ A$_{--}$ and $21_{6,15}-22_{5,18}$ A$_{++}$ at 97.679 GHz, $J_{K_a, K_c}=24_{6,19}-23_{7,16}$ A$_{--}$ and $24_{6,18}-23_{7,17}$ A$_{++}$ at 98.030 GHz, are not detected in this observation. 
Figures 9g and 9h do not well resemble each other although these are the maps in the CH$_3$OH emission lines with $J=2$. 

Figures 9i and 9j show the enlarged integrated intensity maps in the $^{28}$SiO and $^{29}$SiO emission lines with the velocity ranges of $V_{\mathrm{LSR}}=-25$ to $0$ km s$^{-1}$, respectively.  These do not well resemble each other. On the other hand, Figures 9g and 9i resemble each other and Figures 9h and 9j resemble each other in spite of different molecules. The intensity ratio between the object A and the others in Figure 9g is smaller than that in Figure 9h although the intensity in the former is larger than that in the latter. The difference between the two CH$_3$OH maps is similar to that between the maps of the $^{28}$SiO and $^{29}$SiO emission lines, which are the major and minor isotopes (see Figures 9i and 9j).  These would be explained by that the CH$_3$OH (96.741 GHz) and $^{28}$SiO  emission lines are fairly optically thick around the object A but the CH$_3$OH (97.583 GHz) and $^{29}$SiO emission lines are optically thin.  Figure 9h and 9j would indicate faithfully the abundance enhancements of these molecules. Therefore, we conclude that the shocked molecular gas is enhanced strongly in the object A although it is somewhat enhanced in other components. This is consistent with the detection of the Class-I methanol maser only in the object A as mentioned above.

\begin{figure}
\begin{center}
\includegraphics[width=17cm,   bb=0 0  504.14 336.03]{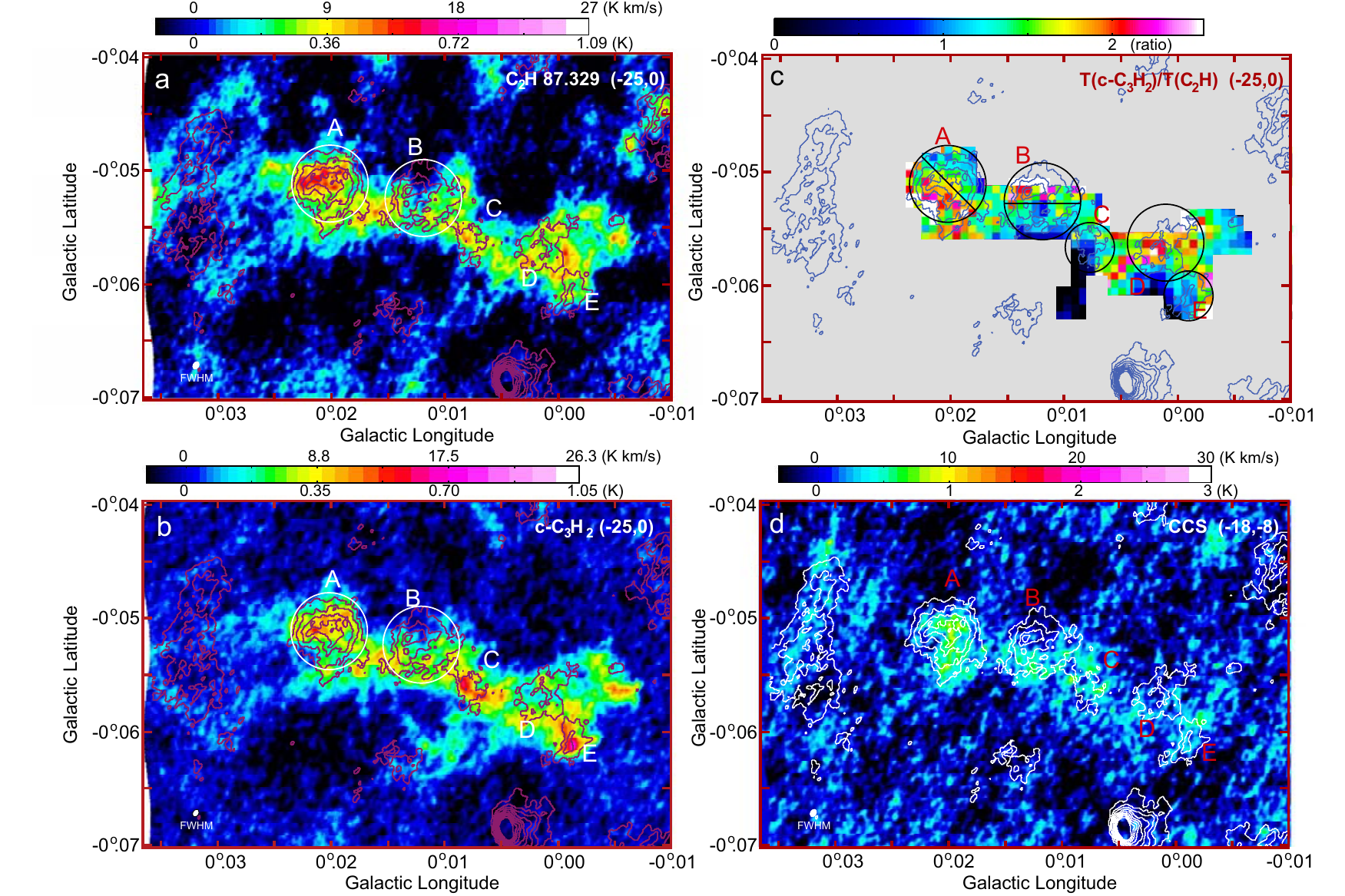}
 \end{center}
 \caption{
 {\bf a} Integrated intensity map of of M0.014-0.054 in the C$_2$H~$N=1-0~J=3/2-1/2~F=1-0$ emission line. The velocity range is $V_{\mathrm{LSR}}=-25$ to $0$ km s$^{-1}$. The unit of the color bar is mean brightness temperature. Contours show the continuum map at 86 GHz. The first contour level and contour interval are both 0.014 K.  {\bf b} Integrated intensity map of M0.014-0.054 in the c-C$_3$H$_2$~$J_{K_a, K_c}=2_{1,2}-1_{0,1}$ emission line. {\bf c} The ratio map of $T_\mathrm{B}$[C$_2$H]/$T_\mathrm{B}$[c-C$_3$H$_2$]  in M0.014-0.054. {\bf d} Integrated intensity map of M0.014-0.054 in the CCS~$N,J=7,6-6,5$ (86.181413 GHz) emission line. The velocity range is $V_{\mathrm{LSR}}=-18$ to $-8$ km s$^{-1}$.  }
 \label{Fig11}
  \clearpage
\end{figure}
\begin{figure}
\begin{center}
\includegraphics[width=17cm,   bb=0 0  672 425]{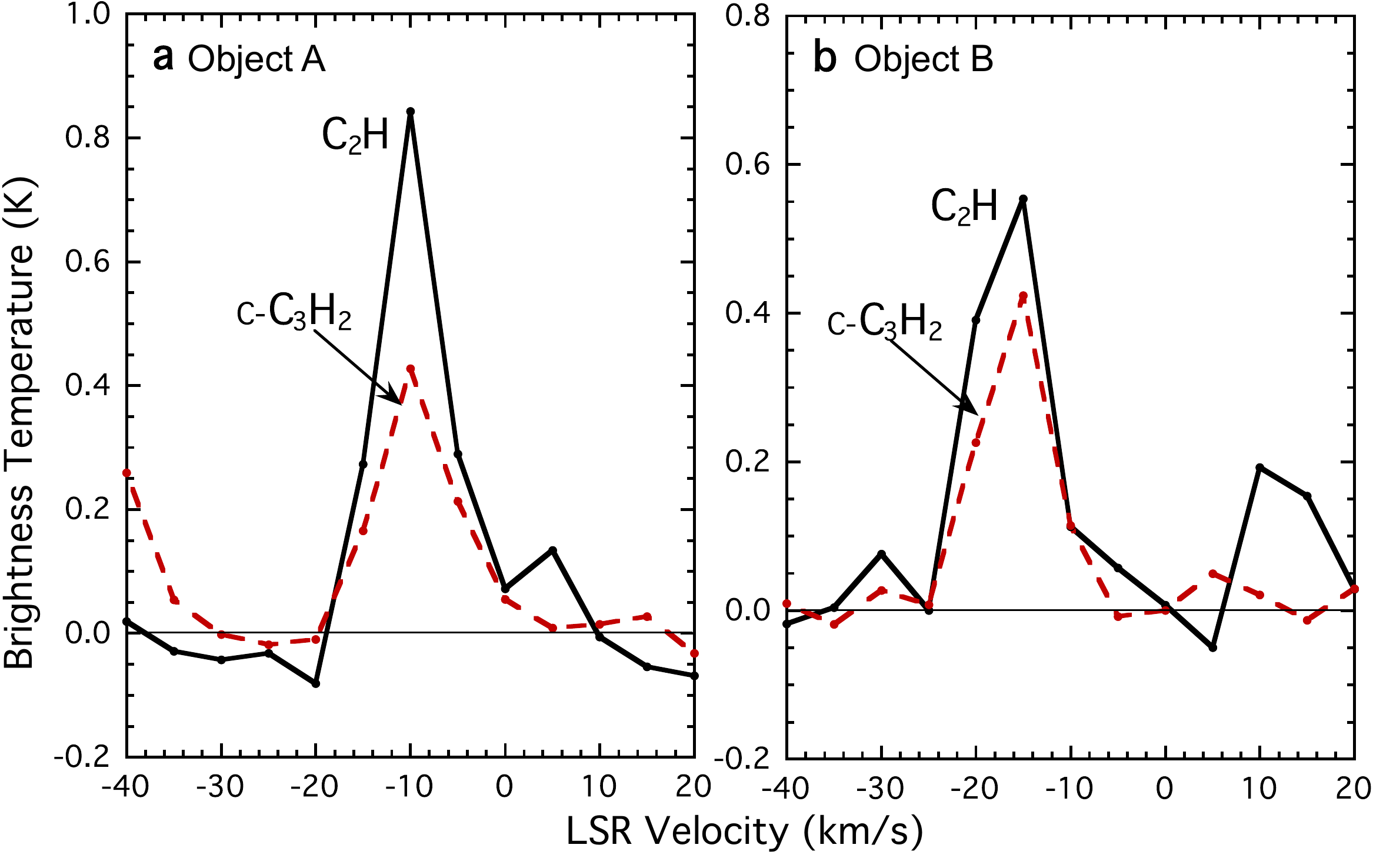}
 \end{center}
 \caption{
 {\bf a} Spectra of the C$_2$H and c-C$_3$H$_2$ $J_{K_a, K_c}=2_{1,2}-1_{0,1}$ emission lines of the object A. The sampling area is shown in Figure 11 (circles). {\bf b} the  C$_2$H and c-C$_3$H$_2$ emission lines of the object B. The sampling area is shown in Figure 11 (circles).} 
 \label{Fig12}
  \clearpage
\end{figure}

\subsection{Embedded Stars in the ``Hot Molecular Core"s}
The C$_2$H and c-C$_3$H$_2$  molecules would fairly survive even in the molecular clouds exposed to UV radiation than others (e.g. \cite{Cuadrado}). 
Figures 11a and 11b show the integrated intensity maps of M0.014-0.054 in the C$_2$H $N=1-0~J=3/2-1/2~F=2-1$ and  c-C$_3$H$_2$ $J_{K_a, K_c}=2_{1,2}-1_{0,1}$ emission lines. The velocity ranges are both  $V_{\mathrm{LSR}}=-25$ to $0$ km s$^{-1}$.  The distributions in these emission lines resembles those in the C$^{34}$S, H$^{13}$CO$^+$ and HN$^{13}$C emission lines. 
The higher $J$ emission line,  c-C$_3$H$_2$ $J_{K_a, K_c}=4_{3,2}-4_{2,3}$, at $85.6564$ GHz is not detected.  Other small hydrocarbon molecules, l-C$_3$H (97.996GHz and 98.012 GHz), C$_4$H (96.478 GHz), and C$_5$H (97.863 GHz) emission lines are not also detected. The transition of c-C$_3$H is not in our observation frequency range. 

Figure 11c shows the ratio map of $T_\mathrm{B}$[C$_2$H]/$T_\mathrm{B}$[c-C$_3$H$_2$] in M0.014-0.054.
In addition, figures 12a and 12b show the spectra of the C$_2$H $N=1-0~J=3/2-1/2~F=2-1$ and  c-C$_3$H$_2$ $J_{K_a, K_c}=2_{1,2}-1_{0,1}$ emission lines  on the objects A and B, respectively. The sampling areas are shown in Figures 11a and 11b as circles. 
The ratio in the object A is $T_\mathrm{B}$[C$_2$H]/$T_\mathrm{B}$[c-C$_3$H$_2$]$=1.73\pm0.03$.
However, the ratio is seen to be inhomogeneous in the object.
The ratio of the southeastern half is $T_\mathrm{B}$[C$_2$H]/$T_\mathrm{B}$[c-C$_3$H$_2$]$=2.06\pm0.05$ (see Figure 11c).  The ratio of the northwestern half is $T_\mathrm{B}$[C$_2$H]/$T_\mathrm{B}$[c-C$_3$H$_2$]$=1.47\pm0.04$ (see Figure 11c). 
On the other hand, the ratio in the object B is $T_\mathrm{B}$[C$_2$H]/$T_\mathrm{B}$[c-C$_3$H$_2$]$=1.53\pm0.02$(see Figure 11c). The ratios of the southern half and northern half are $T_\mathrm{B}$[C$_2$H]/$T_\mathrm{B}$[c-C$_3$H$_2$]$=1.31\pm0.03$ and $=1.96\pm0.05$.  Although the integrated intensities of the emission lines in the northern half are weaker than those in the southern half, the ratio in the northern half is larger than that in the southern half. The ratios in the objects C, D, and E are $T_\mathrm{B}$[C$_2$H]/$T_\mathrm{B}$[c-C$_3$H$_2$]$=1.40\pm0.04$, $1.84\pm0.04$, and  $1.47\pm0.04$, respectively(see Figure 11c).  
The ratios in the objects C, and E are about $\sim1.5$. These are located on the intensity ridge of the molecular cloud in M0.014-0.054 (see Figure 11a and 11b). The ratio of the object D is slightly higher than the others.

Although the photodissociation potential of the C$_2$H molecule, 4.9 eV (C$_2$H $\rightarrow$ C$_2$ + H), is similar to that of the c-C$_3$H$_2$ molecule, 4.4 eV (c-C$_3$H$_2$ $\rightarrow$ C$_3$H + H), the ionization potential of the C$_2$H molecule, 11.4 eV, is fairly higher than that of the c-C$_3$H$_2$ molecule, 9.2 eV (\cite{Hemert}).  The difference between the ionization potentials may raise the viability of C$_2$H molecule in the UV irradiated condition compared with that of the c-C$_3$H$_2$ molecule and increase the abundance ratio of $X$[C$_2$H]/$X$[c-C$_3$H$_2$].
Consequently, $T_\mathrm{B}$[C$_2$H]/$T_\mathrm{B}$[c-C$_3$H$_2$] ratio would increase. 

The high ratios measured in the southeastern half of the object A and the northern half of the object B are distinct from those in the northwestern half of the object A, the southern half of the object B, the object C and the object E. The $T_\mathrm{B}$[C$_2$H]/$T_\mathrm{B}$[c-C$_3$H$_2$] ratios have been reported to be increased in UV irradiated regions, for example, $\sim3$ at the Horse head PDR (\cite{Teyssier}).
The measured high ratios may be consistent with that of the Horse head PDR.  Similar UV irradiated condition is expected in both M0.014-0.054 and the Horse head PDR.
Because the object B is adjacent to the object A, these are expected to be in the similar UV irradiated condition. 
If the UV photons with softer spectrum, for example $h\nu_\mathrm{typ.}\sim4-5$ eV, are dominated in the objects,  a slight difference between the photodissociation potentials of these molecules would decrease the abundance ratio between $X$[C$_2$H] and $X$[c-C$_3$H$_2$].
Consequently, $T_\mathrm{B}$[C$_2$H]/$T_\mathrm{B}$[c-C$_3$H$_2$] ratio is decreased. These suggest that the lower ratios of $\sim1.5$ are mainly determined by the embedded stars or protostars. 

Figure 11d shows the enlarged integrated intensity map of M0.014-0.054 in the CCS~$N,J=7,6-6,5$ (86.181413 GHz) emission line with the velocity range of $V_{\mathrm{LSR}}=-18$ to $-8$ km s$^{-1}$.  
The same velocity range as in other panels cannot be set because this emission line is at the edge of the observation frequency range.  The faint mission is centered in the object A. The emission in the object B is not centered in it but traces the southern limb of the object. The faint emission is probably associated with the objects C, D and E.
The objects A and B are detected in the dust continuum emission but not detected in the recombination line as shown in Subsection 6.1. The star formation activities had started and the surfaces of the embedded stars had been heated at least to several 100 K.
However, the existence of CCS molecules suggests that the chemical evolution by the star formation activity in the object A is still in the early stage (e.g. \cite{Hirahara}).

On the other hand,  the molecular emission lines including the CCS  emission line are not centered in the object B.  The observed features suggest that CCS molecules are still remained in the hollow structure of the molecular cloud around the embedded star.  The star would emit soft UV radiation enough to dissociate the surrounding molecular cloud. 
This means that the chemical evolution in the object B  is advanced comparing with that in the object A.  
If the star has already been in the main sequence stage, it may be a less massive star than B1 because it does not emit vast Lyman continuum emission enough to ionize the surrounding gas.
In addition, if the faint features seen around the objects C, D, and E in the molecular emission lines including the CCS  emission line are the remnants of the cradle molecular clouds, 
the  chemical evolutions in these objects may be more advanced comparing with that in the object B.

\begin{figure}
\begin{center}
\includegraphics[width=12cm, bb=0 0  233.34 330.72]{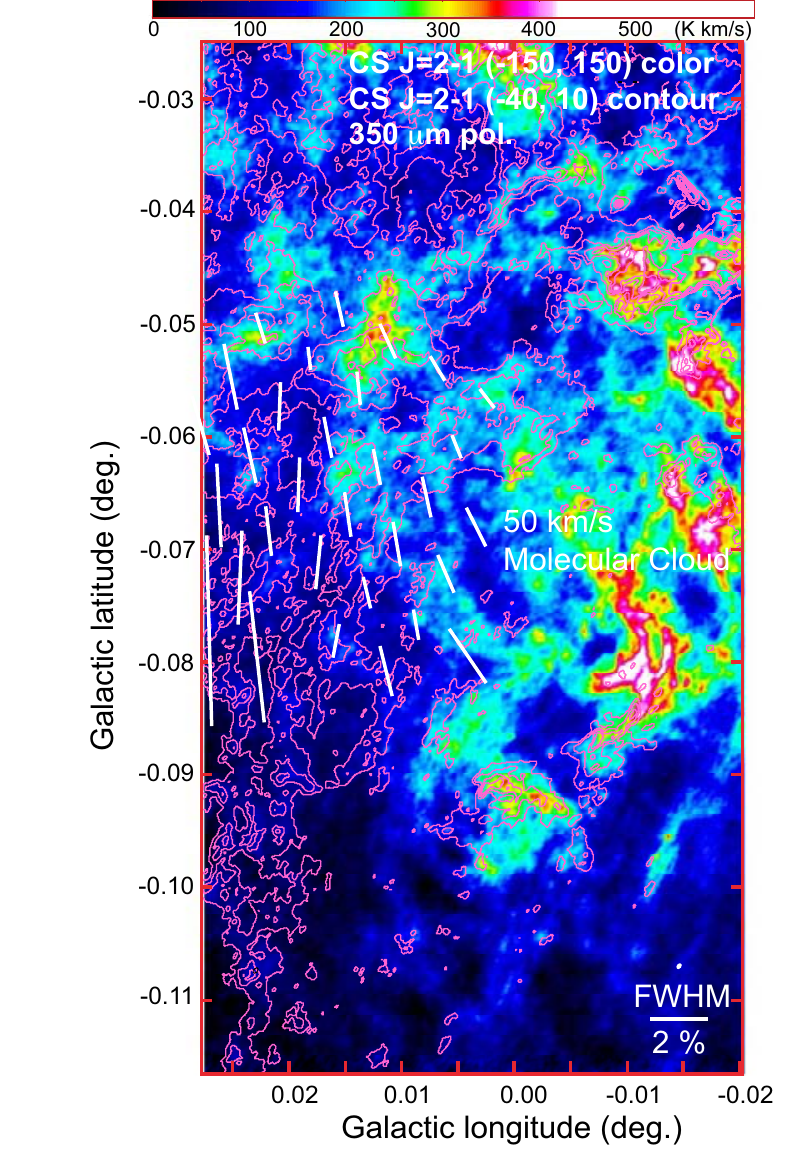}
 \end{center}
 \caption{The vectors indicate the polarization data (direction of $B$ and polarization degree) at $350 \mu$m observed by the Caltech Submillimeter Observatory 10-m telescope (Table 4 in \cite{Chuss2003}), whic are overlaid on integrated intensity maps around the ``Vertical Part" and M0.014-0.054  in the C$^{32}$S $J=2-1$ emission line with the velocity ranges of $V_{\mathrm{LSR}}=-150$ to $150$ km s$^{-1}$ (pseudo color) and $V_{\mathrm{LSR}}=-40$ to $10$ km s$^{-1}$ (contours), respectively. The first contour and contour interval are both $25$ K km s$^{-1}$.
 }
 \label{Fig13}
\end{figure}
\section{Magnetic Field in the ``Vertical Part" }
\subsection{Direction of the Magnetic Field}
We used polarization data at $350 \mu$m with the Caltech Submillimeter Observatory 10-m telescope (CSO) to obtain the direction of the magnetic field, $\phi_{\mathrm{int}}$, in the VP (Table 4 in \cite{Chuss2003}). 
Figure 13 shows the submillimeter polarization ($B$ vector) overlaid on the  velocity integrated intensity map with the velocity ranges of $-150$ to $150$ km s$^{-1}$ (pseudo color) and $-40$ to $10$ km s$^{-1}$ (contours) in the C$^{32}$S $J=2-1$ emission line. There are two molecular cloud components  on the line of sight, of which one belongs to the VP and another is seen in the velocity range of $30$ to $70$ km s$^{-1}$  (see Figures 3b and 4b). 
The observed polarization vectors are the sum of of the intrinsic polarization vectors originated by these components. Although the 50MC is dominant in the western half area, the VP could be identified as the bundle of filaments in the eastern half area. Therefore the observed $B$ vectors in the eastern half area are expected to be originated mainly by the VP. 
We consider that the observed $B$ vectors well trace the magnetic filed lines through the classical Davis-Greenstein mechanism, in which the observed $B$ vectors are thought to be parallel to the magnetic filed lines. The well-ordered $B$ vectors suggest that the magnetic filed lines run along the filaments of the VP. 

Such poloidal magnetic field (or perpendicular to the Galactic plane) has been usually observed in the non-thermal structures  in the Galactic center region (e.g. \cite{Yusef-Zadeh1984}, \cite{Tsuboi1986}). Meanwhile the magnetic field in the Galactic center molecular clouds has been unveiled to be mainly troidal (or parallel to the Galactic plane) by IR observations (e.g. \cite{Nishiyama}).  The poloidal magnetic field in the molecular cloud  is an unique case in the Galactic center region.

\subsection{Strength of the Magnetic Field}
The Chandrasekhar-Fermi method is used to estimate the magnetic field strength from the direction fluctuation of the magnetic filed line and internal gas kinematics (\cite{Chandrasekhar}). The magnetic field strength on the plane of sight is given by
\begin{equation}
\label{ }
B_\bot=\sqrt{4\pi\rho}\frac{\delta v}{\delta \phi},
\end{equation}
where $\rho$, $\delta v$, and $\delta \phi$ are gas density, velocity dispersion, and direction fluctuation of the magnetic field, respectively. This is converted to the following formula for molecular clouds (\cite{Crutcher2004});
\begin{equation}
\label{ }
B_\bot [\mu \mathrm{Gauss}]\simeq9.3\times\sqrt{n\mathrm{(H_2)}}\frac{\Delta v_{\mathrm{int}}}{\Delta \phi_{\mathrm{int}}},
\end{equation}
where $\Delta \phi_{\mathrm{int}}$ [degree] is the direction fluctuation of the magnetic field, the $n\mathrm{(H_2)}$[cm$^{-3}$] is the molecular gas density, which can be estimated by the critical density of the observed molecular line, and $\Delta v_{\mathrm{int}}$[km s$^{-1}$] is the FWHM velocity width of the line profile integrated in the area.
The relation between the observed fluctuation, $\Delta \phi_{\mathrm{obs}}$, and $\Delta \phi_{\mathrm{int}}$, is given by  
\begin{equation}
\label{ }
\Delta \phi_{\mathrm{int}}=\sqrt{\Delta \phi_{\mathrm{obs}}^2- \phi_{\mathrm{err}}^2},
\end{equation}
where $\phi_{\mathrm{err}}$ is the mean error of the observation. The $\Delta \phi_{\mathrm{obs}}$ is calculated from the observed magnetic field directions (\cite{Chuss2003}) as the standard deviation. The intrinsic fluctuation of the direction of the magnetic field is estimated to be $\Delta \phi_{\mathrm{int}}\sim9$ deg in the VP (\cite{Chuss2003}). The FWHM velocity width of the C$^{34}$S emission line is derived to be $\Delta v_{\mathrm{int}}=13.8\pm0.8$ km s$^{-1}$ by Gaussian-fit. The emission line is optically thin in the VP as mentioned previously. The sound velocity of C$^{34}$S molecules is estimated to be  $c_{\mathrm{s, C34S}}=\sqrt{\frac{k_{\mathrm{B}}T_{\mathrm{K}}}{\mu_{\mathrm{C34S}}m_{\mathrm{H}}}} \simeq 0.1$ km s$^{-1}$ at $T_{\mathrm{K}}=80$ K where  $\mu_{\mathrm{C34S}}$ is  the molecular weight of a C$^{34}$S molecule, $\mu_{\mathrm{C34S}}=46$, and $m_{\mathrm{H}}$ is the mass of a Hydrogen atom. The broadening of the FWHM velocity width by the sound velocity is negligible.
The effective critical density of the C$^{34}$S emission line is assumed to be $n\mathrm{(H_2)}\sim5\times10^4$cm$^{-3}$ at $T_{\mathrm{K}}=80$ K. Consequently, the magnetic field strength in the VP is estimated to be $B_\bot\sim3$mGauss. 

The Chandrasekhar-Fermi method is not valid when the magnetic field line is significantly fluctuated by the external perturbations such as HII regions and/or SNRs because the fluctuation of the magnetic field is assumed to be originated only by Alfv\`en wave in the method.  Because the CMZ is generally filled with the external  perturbations, the magnetic field must suffer from such perturbations in varying degrees. The estimated value  of the magnetic field strength should have large ambiguity.

\subsection{Stability of Molecular Cloud Filaments}
The critical mass per unit length of molecular filaments is given by 
\begin{equation}
\label{ }
M_{\mathrm{line, crit}}=\frac{2\sigma_{\mathrm{tot}}^2}{G},
\end{equation}
where $\sigma_{\mathrm{tot}}$ is the total velocity dispersion which is given by
$\sigma_{\mathrm{tot}}=\sqrt{\sigma_{\mathrm{nonth}}^2+c_{\mathrm{s}}^2+\frac{1}{2}V_{\mathrm{A}}^2}$ for a magnetized molecular cloud filament (e.g. \cite{Fiege}, \cite{Arzoumanian}). 
When the LTE mass per unit length of the filament is larger than this limit, the filament can fragment into molecular cores along the axis of the filament (e.g. \cite{inutsuka}). 
In the case of the VP, the velocity dispersion of the nonthermal motion is calculated to be $\sigma_{\mathrm{nonth}}=\sqrt{\frac{\Delta v_{\mathrm{int}}^2-c_{\mathrm{s, C34S}}^{2}}{8ln2}}=5.9\pm0.3$ km s$^{-1}$. The sound velocity in the VP is estimated to be  $c_{\mathrm{s}}=\sqrt{\frac{k_{\mathrm{B}}T_{\mathrm{K}}}{\mu m_{\mathrm{H}}}} \simeq 0.5$ km s$^{-1}$ at $T_{\mathrm{K}}=80$ K where  $\mu$ is  the mean molecular weight, $\mu=2.8$. While the Alfv\`en velocity is estimated to be $V_{\mathrm{A}} = \frac{1300B}{\sqrt{n({\mathrm{H_2}}})}\simeq 17$ km s$^{-1}$ at the magnetic field strength of  $B\sim3$ mGauss and the molecular gas density of $n(\mathrm{H_2})\sim5\times10^4$ cm$^{-3}$ as mentioned in the previous subsection.  
The sound velocity is much smaller than the Alfv\`en velocity and the velocity dispersion of the nonthermal motion.
Therefore, the critical mass per unit length of the VP is estimated to be $M_{\mathrm{line, crit}}=\frac{2\sigma_{\mathrm{tot}}^2}{G} \sim 8\times 10^4$ $M_\odot$ pc$^{-1}$. Because the magnetic field strength is not observed in the MFB, the critical mass per unit length cannot be derived. However, the lower limit can be estimated when the magnetic field is ignored. In the case of the MFB, the velocity dispersion of the nonthermal motion is calculated to be $\sigma_{\mathrm{nonth}}=5.1\pm0.3$ km s$^{-1}$. The critical  mass per unit length of the MFB is $M_{\mathrm{line, crit}}\gtrsim1\times 10^4$ $M_\odot$ pc$^{-1}$.

On the other hands, the LTE mass per unit length of the VP is given by 
\begin{equation}
\label{ }
M_{\mathrm{line, LTE}}\sim\frac{M_{\mathrm{LTE}}}{nL},
\end{equation}
where $M_{\mathrm{LTE}}$ is the LTE molecular cloud mass estimated in Subsection 4.4, $M_{\mathrm{LTE}}\sim4\times10^4(\frac{T_\mathrm{ex}}{80}) M_\odot$, $L$ is the the length of the VP, $L\sim8$ pc, and $n$ is the number of the observed filaments, $n\sim3$ (see Figures 3a and 4a). Therefore, the LTE mass per unit length of the VP is estimated to be
$M_{\mathrm{line, LTE}}\sim2\times10^3(\frac{T_\mathrm{ex}}{80})$ $M_\odot$ pc$^{-1}$. Using the same procedure, the LTE mass per unit length of the MFB is estimated to be $M_{\mathrm{line, LTE}}\sim3\times10^3(\frac{T_\mathrm{ex}}{80})$ $M_\odot$ pc$^{-1}$. 
Because the inclination angles of the VP and MFB are not known, these LTE masses per unit length are the upper limits. 
Both the LTE masses per unit length of the VP and MFB are much smaller than their critical masses per unit length, respectively. The filaments are stable for the fragmentation along the filaments. The star formation activity in the VP and MFB cannot start without any external trigger. 
 In this observation, we found the evidences of the CCC  including the ``Bridge"s between the VP and MFB, and also found the evidences of the star formation including HMCs in M0.014-0.054, which is located at the intersection between the colliding filaments.  Even for stable molecular filaments, the CCC is presumably make a role in the star formation as the external trigger (see also Figure 1 in \cite{Inoue}).
Although there are still many issues in massive star formation, the star formation induced by CCC would be a promising mechanism in the SgrAMC.

\begin{figure}
\begin{center}
\includegraphics[width=17cm, bb=0 0  850 441]{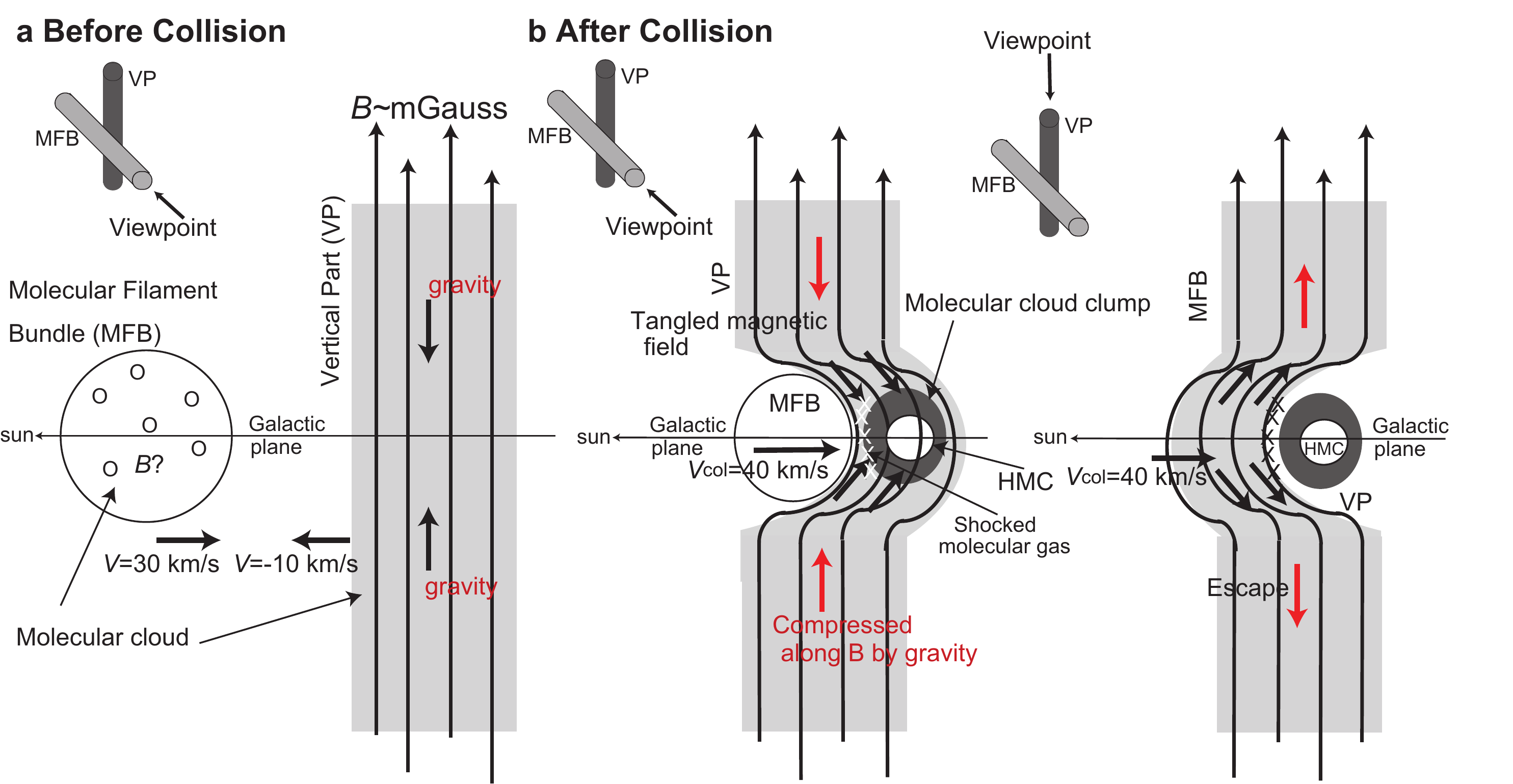}
 \end{center}
 \caption{{\bf a} Schematic display of the VP  and MFB  before the collision between them.
 The viewpoint is shown on the upper-left corner. The magnetic fields in the VP  and MFB are perpendicular to the Galactic plane and parallel to it, respectively. Although the magnetic field in the MFB is not observed, it is expected that this is  parallel to the Galactic plane  because the magnetic field in the neighboring molecular clouds is usually along the Galactic plane  (e.g. \cite{Nishiyama}). The molecular clouds in them are thought to be along the magnetic fields. The molecular clouds in the VP are considered to be confined around the Galactic plane by the gravity although the molecular clouds  in the MFB are considered to be able to move along the magnetic fields.
{\bf b} Schematic display of the VP  and MFB after the collision. 
The viewpoints of the left and right panels are along the MFB and VP, respectively. 
The collision velocity should be larger than $\Delta\gtrsim40 $ km s$^{-1}$ because the SiO molecules seem  to be  abundant  in the colliding area, which are confined  in the tangled magnetic field there. 
In the VP, the molecular clouds should converge kinematically along the magnetic fields and form massive molecular cloud cores around the collision area. Successively, they form high mass protostars in their insides and evolve into HMCs.
On the other hand, the molecular clouds in the MFB can move from the colliding area along the magnetic field because there is no confined molecular cloud like in the VP.  Therefore the molecular clouds around the colliding area would become less dense than those in the VP  and cannot form high mass protostars in their insides. }
 \label{Fig14}
\end{figure}

\section{Spatial Structure of the CCC and It's Possible Scenario}
Based on the observation results and discussion mentioned above, we would like to depict the spatial structure of the CCC between the VP and MFB and infer the possible scenario. As mentioned in the previous section, the VP and MFB, except for the colliding area, are stable molecular filaments against gravitational fragmentation, which are approximately perpendicular to the Galactic plane and parallel to it, respectively. The stabilities are secured by the strong magnetic field and/or large velocity dispersion in the filaments. 

Figure 14a shows the schematic display of the VP and MFB before the collision. The viewpoint of the panel is along the MFB. The directions of the magnetic fields observed by CSO is considered to indicates the large scale structure of the magnetic fields in the VP because the effective angular resolution is as large as $\sqrt{\mathrm{FWHM}^2+\mathrm{Grid}^2}\sim27\arcsec$ (\cite{Chuss2003}). 
The magnetic field in the VP is approximately perpendicular to the Galactic plane (see Figure 13). The molecular filaments depicted in the CS emission line are along the magnetic field. Note that it has not been clear how the vertical magnetic field are originated and how molecular clouds are took into the VP.
On the other hand, those in the MFB are not observed although the molecular filaments are approximately  parallel to the Galactic plane  (see Figure 5). However, because the magnetic field along the Galactic plane is observed in the neighboring molecular clouds (e.g. \cite{Nishiyama}), such magnetic field may  exist also in the MFB.
Molecular clouds can generally move along the magnetic fields even if they have $B\sim$mGauss  in the MFB. 
On the other hand, those in the VP may not be such case. 
As mentioned above,  the molecular clouds  in the VP would be distributed perpendicular to the Galactic plane. It is impossible that they orbit in Kepler motion keeping this distribution. The portions  obeying the Kepler motion don't feel the gravity although the portions departing from it feel the gravity.
If the molecular cloud near the  the Galactic plane moves along the magnetic field,  it must  push up other clouds captured around the Galactic plane by the gravity. Therefore, the molecular clouds in the VP are confined around the Galactic plane by the gravity.  

Figure 14b shows the schematic display of the VP and MFB after the collision. 
The viewpoint of the left panel is along the MFB, which is the same as that in Figure 14a. While, that of the right panel is perpendicular to the Galactic plane, which is along the VP. 
The collision velocity should be larger than $\Delta\gtrsim40 $ km s$^{-1}$ because the SiO emission line is enhanced around the expected colliding area. The molecular filaments  in the VP are seen to get tangled up around M0.014-0.054 although they are lined up perpendicular to the Galactic plane in the farther area from it (see Figures 2 and 3). This probably shows that the magnetic field in the VP got tangled up in the colliding area because the filaments trace the magnetic field.
In the VP, the molecular clouds should converge kinematically along the magnetic fields because the magnetic fields are deformed as shown in Figure 14b (C.f. \cite{Inoue}), and form massive molecular cloud cores around the collision area. Because the shocked molecular gas made by the collision cannot expand crossing the filaments, this should be confined around the colliding area.
The molecular cloud cores were massive enough to be bound gravitationally as mentioned in Subsection 6.2. Therefore, they form high mass protostars in their insides and evolve into HMCs.
On the other hand, the molecular clouds in the MFB can escape from the colliding area along the magnetic field because there is no confined molecular cloud like in the VP as shown the right panel of Figure 14b.  Therefore the molecular clouds around the colliding area would become less dense than those in the VP  and cannot form high mass protostars in their insides. This scenario explains that the HMCs are observed only in the VP.

\begin{ack}  
This work is supported in part by the Grant-in-Aids from the Ministry of Eduction, Sports, Science and Technology (MEXT) of Japan, No.16K05308 and No.19K03939. We are grateful to Prof. S. Takano at Nihon University for useful discussion.
This paper is partly based on observations by the Nobeyama 45-m radio telescope.
The telescope is operated by Nobeyama Radio Observatory, a branch of National Astronomical Observatory of Japan (NAOJ). 
The National Radio Astronomy Observatory (NRAO) is a  facility of the National Science Foundation (NSF) of USA operated under cooperative  agreement by Associated Universities, Inc.. This paper also makes use of the following ALMA data:ADS/JAO.ALMA\#2012.1.00080.S.  ALMA is a partnership of ESO (representing its member states), NSF (USA) and NINS (Japan), together with NRC(Canada), NSC and ASIAA (Taiwan), and KASI (Republic of Korea), in cooperation with the Republic of Chile. The Joint ALMA Observatory is operated by ESO, AUI/NRAO and NAOJ.  This work also has made use of the JCMT Science Archive (http://www.cadc-ccda.hia-iha.nrc-cnrc.gc.ca/en/jcmt/). The JCMT has historically been operated by the Joint Astronomy Centre on behalf of the Science and Technology Facilities Council of the United Kingdom, the NRC, and the Netherlands Organization for Scientific Research.

\end{ack}


\begin{thebibliography}{}
\bibitem[Ao et~al. (2013)]{Ao2013} Ao, Y. et~al. \ 2013, \aap 550, id A135
\bibitem[Armijos-Abenda\~{n}o et~al. (2015)]{Armijos2015} Armijos-Abenda\~{n}o, J., Mart\'{\i}n-Pintado, J., Requena-Torres, M. A., Mart\'{\i}n, S., \& Rodr\'{\i}guez-Franco, A. \ 2015, \mnras, 446, 3842
\bibitem[Arzoumanian et~al. (2013)]{Arzoumanian}Arzoumanian, D., Andr\'e, P., Peretto, N., \& K\"{o}nyves, V. \ 2013, \aap, 553, id A119
\bibitem[Bally et~al. (1987)] {Bally1987}Bally, J., Stark, A. A., Wilson, R. W., \& Henkel, C., \ 1987, \apjs, 65, 13
\bibitem[Boehle et~al. (2016)] {Boehle}Boehle, A. et~al.\ 2016, \apj, 830, 17
\bibitem[Chandrasekhar \&  Fermi (1953)]{Chandrasekhar}Chandrasekhar, S.\&  Fermi, E. \ 1953, \apj, 118, 113
\bibitem[Chuss et~al. (2003)]{Chuss2003}Chuss, D. T., Davidson, J. A., Dotson, J. L., Dowell, C. D., Hildebrand, R. H., Novak, G., \& Vaillancourt, J. E. \ 2003, \apj, 599, 1116
\bibitem[Crutcher et~al. (2004)]{Crutcher2004}Crutcher, R. M., Nutter, D. J., Ward-Thompson, D,, \& Kirk J. M., \ 2004, \apj, 600, 279
\bibitem[Cuadrado et~al. (2015)]{Cuadrado}Cuadrado, S.,  Goicoechea, J. R., Pilleri, P.,  Cernicharo, J., Fuente, A., \&  Joblin, C. \ 2015, \aap, 575, A82
\bibitem[Etxaluze et~al. (2016)] {Etxaluze}Etxaluze, M., Smith, H. A.,  Tolls, V.,  Stark, A. A., \& Gonz\'alez-Alfonso, E., \ 2011, \aj, 142, id 134
\bibitem[Fiege \& Pudritz (2000)]{Fiege}Fiege, J. D., \& Pudritz, R. E. \ 2000, \mnras, 311, 85
\bibitem[Figer et~al. (1999)]{Figer1999} Figer, D.~F.; McLean, I.~ S., \& Morris, M., \ 1999, \apj, 514, 202
\bibitem[Figer et~al. (2002)]{Figer2002} Figer, D.~F.  Najarro, F., Gilmore, D., Morris, et~al.
\ 2002, \apj, 581, 258
\bibitem[Fontani, et~al. (2018)]{Fontani} Fontani, F.,   Vagnoli, A.,  Padovani, M.,  Colzi, L.,  Caselli, P.,  \&   Rivilla, V. M. \ 2018, \mnras, 481, 79
\bibitem[Fuente, Rodriguez-Franco, \& Martin-Pintado (1996)]{Fuente}Fuente, A., Rodriguez-Franco, A., \& Martin-Pintado, J. \ 1996, \aap, 312, 599
\bibitem[Genzel et~al. (1996)]{Genzel}Genzel, R., Thatte, N., Krabbe, A., Kroker, H., \& Tacconi-Garman, L. E. \ 1996, \apj, 472, 153
\bibitem[Gusdorf et~al. (2008)]{Gusdorf}Gusdorf, A.,  Cabrit, S., Flower, D.R.,  \& Pineau des For\^ets, G., \ 2008, \aap, 482, 809
\bibitem[Hasegawa et~al. (1994)] {Hasegawa1994}Hasegawa, T., Sato, F., Whiteoak, J.B., Miyawaki, R., \ 1994, \apjl. 429, L77
\bibitem[Hasegawa et~al. (2008)] {Hasegawa2008}Hasegawa, T., Arai, T., Yamaguchi, N., Sato, F., \ 2008, \apss,  313, 91
\bibitem[Hartquist et~al. (1995)]{Hartquist}Hartquist, T. W., Menten, K. M., Lepp, S., \& Dalgarno, A. \ 1995, \mnras, 272, 184	
\bibitem[Haworth et~al. (2015)]{Haworth}Haworth, T. J., Tasker, E. J., Fukui, Y. et~al., \ 2015, \mnras, 450, 10
\bibitem[Haworth et~al. (2015b)]{Haworthb}Haworth, T. J., Shima, K., Tasker, E. J.  et~al.,
 \ 2015, \mnras, 454, 1634
\bibitem[van Hemert \& van Dishoeck (2008)]{Hemert} van Hemert, M. C. \& van Dishoeck, E. F.  \ 2008, ChemPhys,  343, 292
\bibitem[Hirahara et~al. (1992)]{Hirahara}Hirahara, Y., Suzuki, H., Yamamoto, S., Kawaguchi, K., Kaifu, N., Ohishi, M., Takano, S., Ishikawa, S., \& Masuda, A. \ 1992, \apj, 394, 539
\bibitem[Inoue \&Fukui (2013)]{Inoue} Inoue, T. \& Fukui, Y. \ 2013, \apj, 774, id. L31
\bibitem[Inutsuka \& Miyama (1997)]{inutsuka}Inutsuka, S.-i., \& Miyama, S. M. \ 1997, \apj, 480, 681
\bibitem[Jansen et~al. (1995)]{Jansen} Jansen, D. J., Spaans, M., Hogerheijde, M. R.,\&van Dishoeck, E. F. \ 1995, \aap, 303, 541
\bibitem[Jim\'enez-Serra et~al. (2008)]{Jim} Jim\'enez-Serra, I., Caselli,  Mart\'{\i}n-Pintado, P.J., \&  Hartquist, T. W., \ 2008, \aap 482, 549
\bibitem[LaRosa et~al. (2000)]{LaRosa} LaRosa, T. N., Kassim, Namir E., Lazio, T. Joseph W., \& Hyman, S. D.
\ 2000, \aj, 119, 207
\bibitem[McMullin et~al. (2007)]{McMullin}McMullin, J. P., Waters, B., Schiebel, D., Young, W., \& Golap, K. 2007, Astronomical Data Analysis Software and Systems XVI (ASP Conf. Ser. 376), ed. R. A. Shaw, F. Hill, \& D. J. Bell (San Francisco, CA: ASP), 127 
\bibitem[Mezger \& Henderson (1967)]{Mezger}Mezger, P. G.\& Henderson, A. P. \ 1967, \apj, 147, 471 
\bibitem[Minh et~al. (1991)]{Minh}Minh, Y. C., Brewer, M. K., Irvine, W. M., Friberg, P., \& Johansson, L. E. B. \ 1991, \aap, 244, 470
\bibitem[Miyawaki et~al. (2020)]{Miyawaki}Miyawaki, R., Tsuboi, M., Kitamura, Y., Uehara, K.,\& Miyazaki, A., \ 2020, in preparation.
\bibitem[Morris (1993)] {Morris1993} Morris, M. R., \ 1993, \apj, 408, 496
\bibitem[Morris \& Serabyn(1996)]{MorrisSerabyn} Morris, M.,\& Serabyn, E. \ 1996, AAR\&A, 34, 645
\bibitem[Nagy et~al. (2015)]{Nagy}Nagy, Z., Ossenkopf, V., Van der Tak, F. F. S., et~al. \ 2015, \aap, 578, A124
\bibitem[Nishiyama et~al. (2010)]{Nishiyama}Nishiyama, S., Hatano, H., Tamura, M., et~al. \ 2010, \apj, 722, L23
\bibitem[Oka et~al. (1998)] {Oka1998} Oka, T., Hasegawa, T., Sato, F., Tsuboi, M., \& Miyazaki, A., \ 1998, \apjs, 118, 455
\bibitem[Oka et~al. (2019)] {OkaT} Oka, T., Geballe, T. R., Goto, M., Usuda, T., Benjamin, J. M., \& Indriolo, N. \ 2019, \apj, 883, id 54
\bibitem[Pierce-Price et~al. (2000)]{Pierce-Price2000} Pierce-Price, D. et~al., \ 2000, \apjl, 545,  L121
\bibitem[Sakai et~al. (2008)]{Sakai}Sakai, N., akai, T., Aikawa, Y., Yamamoto, S. \ 2008, \apjl, 675, L89
\bibitem[Serabyn \& G\"usten (1987)]{Serabyn} Serabyn, E. \& G\"usten, R. \  1987, \aap, 184,133
\bibitem[Schmidt, Zack, \& Ziurys (2018)]{Schmidt}Schmidt, D. R., Zack, L. N. \& Ziurys, L. M. \ 2018, \apjl, 864, id. L31
\bibitem[St\'ephan et~al. (2018)]{Stephan}St\'ephan, G., S., Schilke, P., Le Bourlot, J., Schmiedeke, A., Choudhury, R., Godard, B., \& S\'anchez-Monge, \'A. \ 2018, \aap, 617, 60
\bibitem[Teyssier et~al. (2004)]{Teyssier}Teyssier, D., Foss\'e, D., Gerin, M., Pety, J., Abergel, A.\& Roueff, E. \ 2004, \aap, 417, 135
\bibitem[Tielens (2013)]{Tielens}Tielens, A. G. G. M., \ 2013, Rev. Mod. Phys., 85, 1021
\bibitem[Tsuboi et~al. (1986)]{Tsuboi1986} Tsuboi, M., Inoue, M., Handa, T., Tabara, H., Kato, T., Sofue, Y., \& Kaifu, N.    \ 1986, \aj, 92, 818
\bibitem[Tsuboi, Handa \& Ukita (1999)]{Tsuboi1999} Tsuboi, M., Handa, T., \& Ukita, N. \ 1999, \aj, 120, 1
\bibitem[Tsuboi et~al. (2011)]{Tsuboi2011} Tsuboi, M., Tadaki, K-I., Miyazaki, A., \& Handa, T. \ 2011, \pasj, 63, 763
\bibitem[Tsuboi et~al. (2015)]{Tsuboi2015} Tsuboi, M., Miyazaki, A., \& Uehara, K. \ 2015, \pasj, 67, id. 109
\bibitem[Tsuboi et~al. (2019)]{Tsuboi2019}  Tsuboi, M., Kitamura, Y., Uehara, K., Miyazaki, A., Miyawaki, R., Tsutsumi, T., \& Miyoshi, M. \ 2019, \pasj, 71, id.128

\bibitem[Uehara et~al. (2019)]{Uehara}Uehara, K., Tsuboi, M., Kitamura, Y., Miyawaki, R.,  \& Miyazaki, A.,   \ 2019, \apj, 872, id. 121
\bibitem[Uehara et~al. (2020)]{Ueharab}Uehara, K., Tsuboi, M., Kitamura, Y., Miyawaki, R.,  \& Miyazaki, A.,   \ 2020, in preparation.
\bibitem[Yusef-Zadeh,  Morris, \& Chance (1984)]{Yusef-Zadeh1984} Yusef-Zadeh, F., Morris, M.,  \& Chance, D. \ 1984, \nat, 310, 557
\bibitem[Yusef-Zadeh et~al. (2013)]{Yusef-Zadeh2013}Yusef-Zadeh, F., Cotton, W., Viti, S., Wardle, M., \& Royster, M., \ 2013, \apjl, 764, id. L19
\end{thebibliography}
\end{document}